\documentclass[twocolumn,prd,10pt,showpacs,superscriptaddress]{revtex4-1}
\usepackage[utf8]{inputenc}
\usepackage{graphicx}
\usepackage{mathrsfs}
\usepackage{amsmath}
\usepackage{amssymb}
\usepackage{bm}
\usepackage{hyperref}
\usepackage{soul}
\usepackage{color}
\usepackage{braket}
\usepackage{dsfont}
\usepackage{cleveref}
\newcommand{\pt}{\partial}
\newcommand{\mb}{\mathbf}

\newcommand{\egap}{\varepsilon_{\mbox{\scriptsize{gap}}}}

\newcommand{\mc}{\mathcal}
\newcommand{\wext}{\omega_{\mbox{\tiny{ext}}}}
\newcommand{\Text}{T_{\mbox{\tiny{ext}}}}
\newcommand{\gex}{g^{ji}_{\mbox{\tiny{ext}}}}

\newcommand{\chid}{\chi^{\dagger}}
\newcommand{\Scl}{\mathcal{S}_{\mbox{\scriptsize{cl}}}}

\newcommand{\SQ}{\mathcal{S}_{\mbox{\scriptsize{Q}}}}
\newcommand{\Sef}{\mathcal{S}_{\mbox{\scriptsize{eff}}}}
\newcommand{\Sst}{\mathcal{S}_{\mbox{\scriptsize{st}}}}
\newcommand{\Sdis}{\mathcal{S}_{\mbox{\scriptsize{dis}}}}

\newcommand{\Pfr}{\Psi^{\scalebox{0.7}{free}}}

\newcommand{\rcl}{\rangle}
\newcommand{\phiext}{\varphi_{\mbox{\tiny{ext}}}}

\DeclareMathOperator{\Tr}{Tr}

\begin{document}

\title{Quantum Brownian Motion of a Magnetic Skyrmion}

\author{Christina Psaroudaki}
\affiliation{Department of Physics, University of Basel, Klingelbergstrasse 82, 4056 Basel, Switzerland}
\author{Pavel Aseev}
\affiliation{Department of Physics, University of Basel, Klingelbergstrasse 82, 4056 Basel, Switzerland}
\author{Daniel Loss}
\affiliation{Department of Physics, University of Basel, Klingelbergstrasse 82, 4056 Basel, Switzerland}

\date{\today}
%%%%%%%%%%%%%%%%%%%%%%%%%%%%%%%%%%%%%%%%%%%%%%%%%%%%%%%%%%%%%%%%%%%%%%%%
%%                           Abstract
%%
%%%%%%%%%%%%%%%%%%%%%%%%%%%%%%%%%%%%%%%%%%%%%%%%%%%%%%%%%%%%%%%%%%%%%%%%
\begin{abstract}
Within a microscopic theory, we study the quantum Brownian motion of a skyrmion in a magnetic insulator coupled to a bath of  magnon-like quantum excitations. The intrinsic skyrmion-bath coupling gives rise to damping terms for the skyrmion center-of-mass, which remain finite down to zero temperature due to the quantum nature of the magnon bath. We show that the quantum version of the fluctuation-dissipation theorem acquires a non-trivial temperature dependence. As a consequence, the skyrmion mean square displacement is finite at zero temperature and has a fast thermal activation that scales quadratically with temperature, contrary to the linear increase predicted by the classical phenomenological theory. The effects of an external oscillating drive which couples directly on the magnon bath are investigated. We generalize the standard quantum theory of dissipation and we show explicitly that additional time-dependent dissipation terms are generated by the external drive. From these we emphasize a friction and a topological charge renormalization term, which are absent in the static limit. The skyrmion response function inherits the time periodicity of the driving field and it is thus enhanced and lowered over a driving cycle. Finally, we  provide a generalized version of the nonequilibrium fluctuation-dissipation theorem valid for  weakly driven baths. 
\end{abstract}

\maketitle
%%%%%%%%%%%%%%%%%%%%%%%%%%%%%%%%%%%%%%%%%%%%%%%%%%%%%%%%%%%%%%%%%%%%%%%%
%%                           Introduction
%%
%%%%%%%%%%%%%%%%%%%%%%%%%%%%%%%%%%%%%%%%%%%%%%%%%%%%%%%%%%%%%%%%%%%%%%%%
\section{Introduction} 

The impact of the bath fluctuations on the dynamics of open nonequilibrium systems is commonly treated by nonlinear stochastic differential equations for the macrovariables, known as generalized Langevin equations \cite{WeissBook}. Within this description, the thermal bath exerts random fluctuating forces on the central system which eventually undergoes a Brownian propagation \cite{KeizerBook,ReichlBook,Hanggi05}. The system-bath coupling gives rise to non-Markovian memory damping terms and random forces with a colored correlation \cite{LindenbergBook}. In principle, both the noise and the damping terms are determined by the system-bath interaction, a relation which is manifested in the well known fluctuation-dissipation theorem \cite{Brown63}. 

Quantum stochastic dynamics are present in a variety of physical systems, ranging from quantum optics \cite{TannoudjiBook}, transport processes in Josephson junctions \cite{Makhlin01}, coherence effects and macroscopic quantum tunnelling in condensed matter physics \cite{CaldeiraBook} and many more, which form a large body of current active research. Here we focus on the stochastic dynamics of particle-like magnetic skyrmions, which similar to particle-like solitonic textures in quantum superfluids \cite{Efimkin16,Hurst17}, experience dissipative and stochastic forces from their environmental surroundings. 

Skyrmions are spatially localized two dimensional (2D) magnetic textures characterized by a topologically nontrivial charge $Q_0$ \cite{Wilczek83,Papanicolaou91} given by
\begin{align}
Q_0=\frac{1}{4 \pi} \int d \mb{r}~ \mb{m} \cdot (\pt_x \mb{m} \times \pt_y \mb{m}) \,,
\end{align}
where $\mb{m}$ is the normalized magnetization vector field and $x$ and $y$ are the spatial coordinates of the 2D magnetic layer. Besides their early theoretical prediction \cite{Bogdanov89,Roessler11}, magnetic skyrmions have been observed in bulk metallic magnets \cite{Muehlbauer09,Yu10,Park14}, multiferroic insulators \cite{Seki12,White12} as well as ultrathin metal films on heavy-element substrates \cite{Heinze11,Romming13}. Because of their protected topology, nanoscale size, high mobility \cite{Fert13,Jonietz10,Yu12,Sampaio13,Nagaosa13} and controllable creation \cite{Romming13}, they are in the focus of current research as attractive candidates for future spintronic devices \cite{Wiesendanger16}. 

Classically, the dynamics of a magnetic skyrmion is governed by the Landau-Lifshitz- Gilbert (LLG) equation \cite{lifshitzBK80,gilbertTM04}, which incorporates dissipation mechanisms by a phenomenological local in time Ohmic friction term, known as Gilbert damping. At finite temperatures, the skyrmion is subjected to thermal fluctuations that will render its propagation stochastic, similarly to the Brownian motion of a particle. The conventional assumption for the fluctuating field acting on magnetic particles \cite{GPalacios} as well as skyrmions \cite{Troncoso14,Troncoso14b,Schutte14,Barker16,Diaz17,Miltat18,Nozaki19}, is that it is a Gaussian stochastic process with a white noise correlation function proportional to the phenomenological  Gilbert damping. 

In a magnetic insulator and at low enough temperature, the skyrmion dynamics is dominated by the unavoidable coupling of its center-of-mass with the magnetic excitations generated by the skyrmion motion itself. Magnetic excitations are defined as fluctuations around the classical skyrmion solution through a consistent separation between collective (center-of-mass) and intrinsic (magnetic excitations) degrees of freedom. A description of the dynamics of one-dimensional (1D) domain walls \cite{Braun96} and 2D magnetic skyrmions \cite{Psaroudaki17} in a magnetic insulator beyond the classical framework, demonstrated that the dissipation arising from the magnetic excitations is generally non-Markovian with a damping kernel that is nonlocal in time. The quantum nature of the magnetic bath, naturally incorporated within this approach, becomes evident in the nontrivial temperature $T$ dependence of the damping kernel which remains finite even for vanishingly small $T$. A theory of dissipation which ignores quantum effects based on the classical phenomenological LLG equation is expected to be inadequate for atomic-size skyrmions observed in state-of-the-art experiments carried out at low temperatures of a few K\cite{Heinze11,Yu11,Grenz17}.

In this paper we develop a microscopic description of the skyrmion stochastic dynamics at finite temperature using the functional Keldysh formalism for dissipative quantum systems \cite{Grabert88,Keldysh64,Kamenev09}, as well as the Faddeev-Popov collective coordinate approach\cite{SakitaBook,Braun96} to promote the skyrmion center-of-mass to a dynamic quantity. We then arrive at a Langevin equation of motion, which includes a non-Markovian damping kernel and a stochastic field with a colored autocorrelation function, as a result of the skyrmion-magnon bath coupling.  We demonstrate that the quantum version of the fluctuation-dissipation theorem acquires a non-trivial temperature dependence. As an important consequence, the skyrmion mean square displacement is finite at $T = 0$, and has a fast thermal activation being proportional to $T^2$ for finite temperatures, in contrast to the linear $T$-increase obtained within the usual phenomenological theory \cite{Miltat18}.

We also investigate the effects of an external oscillating drive which unavoidably couples with the magnon bath in an analogous fashion to many physical situations where the driving of the bath results in important contributions to the dynamical response of the entire nanoscale system \cite{Grabert15,Frey16,Reichert16,Grabert16}. We demonstrate explicitly that additional time-periodic dissipative terms are generated by the driving field, in particular a friction and a topological charge renormalization term, which are both absent in the static limit. As a consequence, the skyrmion response function inherits the time periodicity of the drive, and it is thus enhanced and lowered over a driving cycle. Since the magnetic excitations are driven out of equilibrium, a generalization of the fluctuation-dissipation theorem should not be expected in general. Quite remarkably, however, in the weak driving regime, we find a nonequilibrium fluctuation-dissipation relation, which reduces to the equilibrium one in the static limit.

For the efficient manipulation of skyrmions at the nanoscale it is important to understand how random processes contribute to the skyrmion propagation, especially in the presence of time-periodic microwave fields which appear to be among the most efficient ways to induce translational motion of skyrmions in magnetic insulators \cite{Wang15,Moon16,Psaroudaki18}. The microscopic understanding of the  stochastic skyrmion motion becomes also important in view of proposed devices for stochastic computing based on skyrmions \cite{Pinna18,Zazvorka18}.

The structure of the paper is as follows. In Sec.~\ref{sec:Model} we present a detailed derivation of the Langevin equation for the skyrmion collective coordinate using the functional Keldysh formalism in the presence of a time-dependent magnetic field. In Sec.~\ref{sec:Dissipation} we evaluate and discuss the damping kernel, while in Sec.~\ref{sec:Response} we investigate the skyrmion response function. The quantum fluctuation-dissipation theorem and its generalized nonequilibrium version in the presence of the oscillating field are presented in Sec.~\ref{sec:FDT}, together with a discussion on the skyrmion mean square displacement. Our main conclusions are summarized in Sec.~\ref{sec:Conclusions}, while some technical details are deferred to four Appendices. 
%%%%%%%%%%%%%%%%%%%%%%%%%%%%%%%%%%%%%%%%%%%%%%%%%%%%%%%%%%%%%%%%%%%%%%%%
%%                           Langevin Equation
%%
%%%%%%%%%%%%%%%%%%%%%%%%%%%%%%%%%%%%%%%%%%%%%%%%%%%%%%%%%%%%%%%%%%%%%%%%
\section{Langevin Equation	}\label{sec:Model} 
The purpose of this section is to present a derivation of the quantum Langevin equation for the skyrmion center-of-mass coordinate, by making use of a functional integral approach for the magnetic degrees of freedom at finite but low temperatures, combined with the Keldysh technique to include the effects of a time-dependent oscillating magnetic field. To begin with, we note that the essential features of the dynamics of a normalized magnetization field in spherical parametrization $\mb{m} = [\sin \Theta \cos\Phi, \sin \Theta \sin \Phi, \cos \Theta ]$ defined in the 2D space, are described by a partition function of the form $Z=\int \mc{D} \Phi \mc{D} \Pi e^{i \mc{S}}$. Here, the functional integration is over all configurations and the field $\Pi= \cos \Theta$ is canonically conjugate to $\Phi$. The Euclidean action $\mc{S}$ for a thin magnetic insulator in physical units of space $\tilde{\mb{r}}$ and time $\tilde{t}$ is given by 
\begin{align}
\mc{S}= \int d\tilde{t} d\tilde{\mb{r}}~[\frac{S N_A}{\alpha^2} \dot{\Phi} (\Pi-1) - N_A \mc{W}(\Phi,\Pi)] \,,
\label{Action}
\end{align}
where $\dot{\Phi} = \pt_{\tilde{t}} \Phi$ denotes the real-time derivative of field $\Phi$. The first term in Eq.~\eqref{Action} describes the dynamics and is known as the Wess-Zumino or Berry phase term \cite{Braun96}, while the translationally symmetric energy term, 
%$\mc{W}$ 
\begin{align}
\mc{W}(\mb{m})= J \left( \nabla_{\tilde{\mb{r}}} \mb{m}  \right)^2 +\frac{D}{\alpha} \mb{m} \cdot \nabla_{\tilde{\mb{r}}} \times \mb{m} - \frac{K}{\alpha^2} m_z^2- \frac{g\mu_B H}{\alpha^2} m_z \,,
\label{FreeEnergy}
\end{align}
supports skyrmion configurations with nontrivial topological number $Q_0$ as metastable solutions due to the presence of the Dzyaloshinskii-Moriya (DM) interaction\cite{Dzyaloshinsky58,Moriya60}  of strength $D$. Here, $\tilde{\mb{r}} = (\tilde{x},\tilde{y})$, $S$ is the magnitude of the spin, $N_A$ is the number of magnetic layers along the perpendicular $\tilde{z}$ axis and $\alpha$ is the lattice spacing. The strength of the exchange interaction $J$, the easy axis anisotropy $K$, and finally $D$ are measured in units of energy while the strength of the magnetic field $H$ is given in units of Tesla (T). 
%%%%%%%%%%%%%%%%%%%%%%%%%%%%%%%%%%%%%%%%%%%%%%%%%%%
\begin{figure}[t]
\centering
\includegraphics[width=1\linewidth]{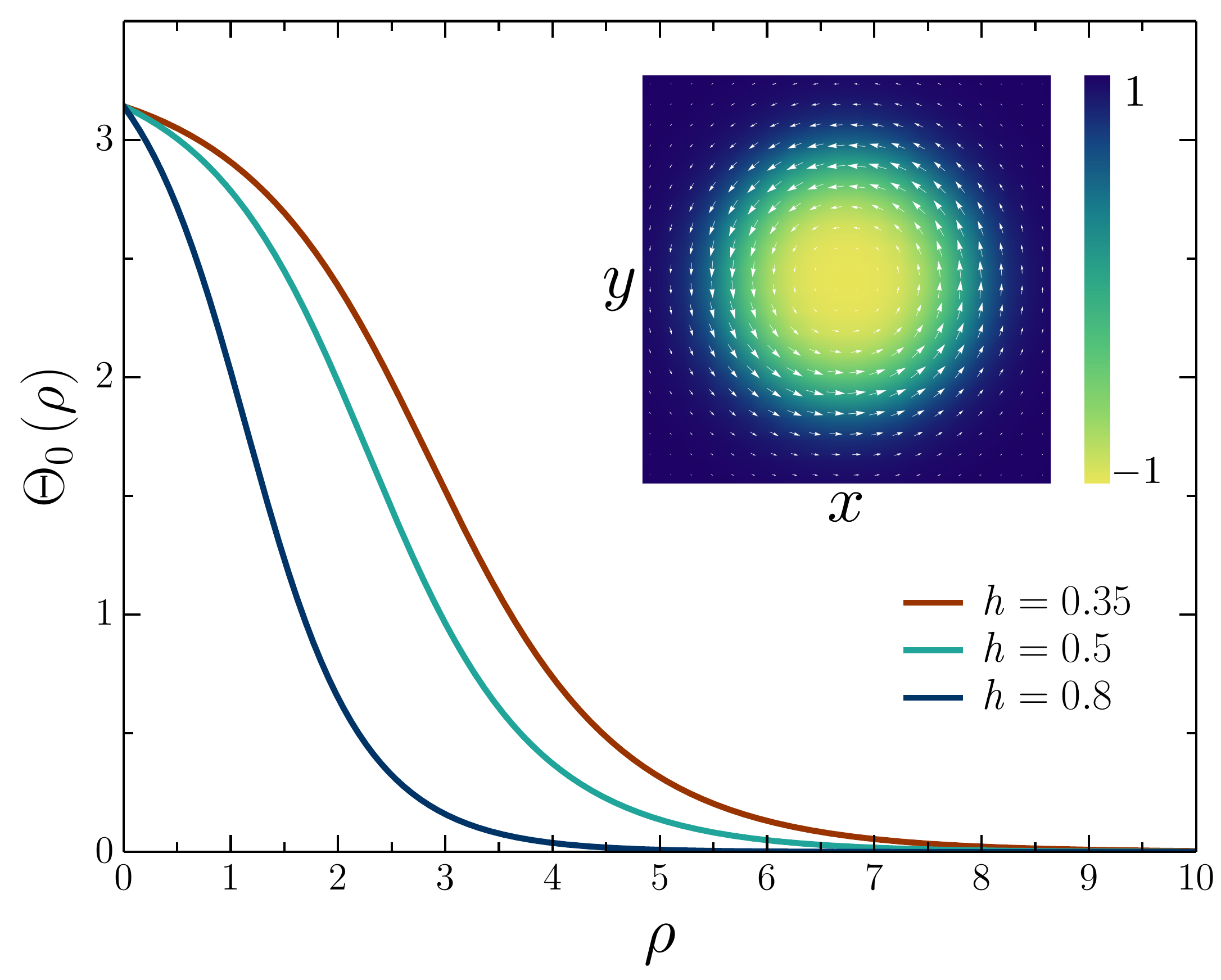}
\caption{\ Magnetization profiles $\Theta_0(\rho)$ of a skyrmion as function of radial distance $\rho$ for  $\kappa=0.1$ and three values of the magnetic field. The colored surface in the inset represents the out-of-plane component of the magnetization texture $\cos \Theta_0$ of a skyrmion with $Q=-1$ in the 2D $xy$ plane for $\kappa=0.1$ and $h=0.35$. 
} \label{Profile}
\end{figure}
%%%%%%%%%%%%%%%%%%%%%%%%%%%%%%%%%%%%%%%%%%%%%%%%%%%

It is convenient to introduce dimensionless variables as $\mb{r}= (D/J \alpha) \tilde{\mb{r}}$, $t= D^2 \tilde{t}/J$, and $T= k_B \tilde{T} J/D^2 $, where $\tilde{T}$ is the temperature measured in Kelvin (K). Also, $k_B$ is the Boltzmann constant and throughout this work we use $\hbar =1$. The energy functional in reduced units is given by
\begin{align}
\mc{F}(\mb{m})= \left( \nabla_{\mb{r}} \mb{m}  \right)^2 + \mb{m} \cdot \nabla_{\mb{r}} \times \mb{m} -\kappa m_z^2 - h m_z \,,
\end{align}
where $\kappa=J K/D^2$, $h=J g\mu_B H/D^2$, and $\mc{F}(\mb{m})=J(\alpha)^2/D^2 \mc{W}(\mb{m})$. The classical skyrmion field, denoted as $\Phi_0(\mb{r})$ and $\Pi_0(\mb{r})$, is found by minimizing the energy functional $\mc{F}(\mb{m})$ \cite{Bogdanov94,Butenko10}. We then arrive at the following rotationally symmetric solution in polar coordinates $\mb{r}=(\rho \cos \phi, \rho \sin \phi)$ given by $\Phi_0(\mb{r})= \phi + \pi/2$, while the skyrmion profile depends only on the radial coordinate $\Theta_0(\mb{r}) = \Theta(\rho)$. In Fig.~\ref{Profile} we depict the magnetization profile of the skyrmion $\Theta_0(\rho)$ for various values of the magnetic field $h$, using the trial function $\Theta_0(\rho) =  A \cos^{-1}(\tanh [(\rho - \lambda)/\Delta_0])$, where $A=\pi/\cos^{-1}(\tanh[-\lambda/\Delta_0])$. The parameter $\lambda$, which denotes the skyrmion size, and $\Delta_0$ are calculated by fitting the approximate function to the one obtained numerically. This profile has a topological number $Q_0=-1$. 

We next address the stochastic dynamics of the skyrmion described by the classical fields $\Phi_0$ and $\Theta_0$ in contact with the bath of magnetic excitations at finite temperature driven by an external  magnetic field that oscillates in time. This is achieved by first promoting the skyrmion center-of-mass to a dynamical variable $\mb{R}(t)$, then treating the magnetic excitations as quantum fluctuations around the classical field, and finally obtaining an effective functional \cite{Braun96,Alamoudi98,WeissBook,Psaroudaki17} by integrating out the magnon degrees of freedom. At the same time, the real-time dynamics of the external field as well as the stochastic effects of the magnon bath at finite $T$ are captured by replacing the time integration by an integration over the Keldysh contour which consists of two branches. The upper branch extends from $t =-\infty$ to $t = +\infty$, while the lower branch extends backwards from $t =\infty$ to $t = -\infty$ \cite{Kamenev09}. It is worth mentioning that the formalism derived below is applicable to any general energy functional $\mc{F}$ as long as it satisfies the specified requirements. 

We define two components of the fields as $\Phi_+ \equiv \Phi(t+i0)$ and $\Phi_- \equiv \Phi(t-i0)$, that reside on the upper and the lower parts of the time contour, respectively. Similarly, we define fields $\Pi_{\pm}=\Pi(t \pm i 0)$. Moreover, quantization of the path integral variables implies the following form:
\begin{align}
\Phi_{\pm}(\mb{r},t) &= \Phi_0^{\pm}(\mb{r}-\mb{R}_{\pm}(t)) + \varphi_{\pm}(\mb{r}-\mb{R}_{\pm}(t),t) \nonumber \\
\Pi_{\pm}(\mb{r},t)&= \Pi_0^{\pm}(\mb{r}-\mb{R}_{\pm}(t)) + \eta_{\pm}(\mb{r}-\mb{R}_{\pm}(t),t) \,,
\label{FullForm}
\end{align}
where $\eta$ and $\varphi$ are the quantum fluctuations and the coordinate $\mb{R}(t)$ is energy independent owing to the assumed translational invariance of the system. We therefore expect the existence of a pair of zero modes $\mc{Y}_i$, with $i=x,y$, which need to be excluded from the functional integral to avoid overcounting degrees of freedom by imposing  proper gauge fixing conditions. We use the following convenient spinor notation,
\begin{equation}
\chi_{\pm} =\frac{1}{2} \binom{\varphi_{\pm} \sin \Theta_0+i \eta_{\pm}/ \sin\Theta_0 }{\varphi_{\pm} \sin \Theta_0-i \eta_{\pm}/ \sin\Theta_0} \,,
\label{SpinorNotation}
\end{equation}
and we also define linear transformations of the fields by performing a Keldysh rotation of the form $\chi_{c,q} = (\chi_+ \pm \chi_-)/\sqrt{2}$ as well as $\mb{R}_{c,q} = (\mb{R}_+ \pm \mb{R}_-)/\sqrt{2}$. Here, $\chi_{c} ~(\mb{R}_{c})$ and $\chi_{q}~(\mb{R}_{q})$ denote the classical and quantum fluctuations (coordinate), respectively. Moreover, we introduce the field $\zeta = \binom{\chi_c}{ \chi_q}$ in order to obtain the action in a more compact form. Implementing all the above transformations  in the action of Eq.~\eqref{Action}, taking into account that time integration is now performed over the upper and lower time branches denoted by the symbol $s=\pm1$, the partition function becomes $Z=\int \mc{D} \mb{R}_c \mc{D} \mb{R}_q e^{i\Scl} \tilde{Z}$, where 
\begin{align}
\tilde{Z} =\int \mc{D} \zeta^{\dagger} \mc{D} \zeta \prod_{s=\pm 1} \delta(F_x^s)\delta(F_y^s) \det(J_{\mb{F}^s}) e^{i \SQ}  \,.
\label{PartFunQ}
\end{align}
Here, $F_i ^s =\int d\mb{r} \chid_s \sigma_z \mc{Y}_i$ is the gauge condition and  $J_{\mb{F}}^s (t,t')= d \mb{F}^s(t)/d\mb{R}(t')$ is the Jacobian matrix of the coordinate transformation and is treated as additional perturbation to the $N_A$-term in the action. The classical part of the effective action reads 
\begin{align} 
\Scl=N_A d \int_{t,\mb{r}}\sum_{s=\pm1}s [-S \dot{\mb{R}}_s \Pi_0^{s} \nabla \Phi_0^{s} -\mb{b}\cdot \mb{m}(\Phi_0^{s},\Pi_0^s) ]  \,,
\label{ActionCl}
\end{align}
where $\mb{b}(t)$ denotes a time-dependent external field, $d=(J/D)^2$ and we have also neglected an overall constant from the configuration energy of the classical skyrmion $\mc{S}_0 =d \int_{\mb{r},t} \mc{F}(\Phi_0,\Pi_0)$. The fluctuation-dependent part of the Keldysh action takes the form 
\begin{align}
\SQ=N_A d ~\zeta^{\dagger} \circ \hat{G}^{-1} \zeta  \,,
\label{ActionQ}
\end{align}
where $\hat{G}^{-1} =(\mc{G}_0^{-1}-V +\frac{1}{\sqrt{2}} \mc{K}_c ) \sigma_x + \frac{1}{\sqrt{2}} \mc{K}_q \mathds{1}$. The magnon Green function is $\mc{G}_0^{-1}=i S\sigma_z \partial_{t} - \mathcal{H}$ and the Hamiltonian  is defined as $\mathcal H=\delta_{\chi^\dagger}\delta_\chi\mathcal F |_{\chi=\chi^\dagger=0}$. The potential $V(\mb{r},t) = \mb{b}(t) \cdot \mb{D}$ with $ \mb{D} =  \delta_{\chi^\dagger}\delta_\chi \mb{m} |_{\chi=\chi^\dagger=0}$ describes the coupling of the external field with the magnons and it is treated as a time-dependent perturbation to the magnon Hamiltonian. The magnetic fluctuations appear as solutions of the eigenvalue problem (EVP) $\mc{H} \Psi_{n} = \varepsilon_{n} \sigma_z \Psi_{n}$, solved in detail in Appendix~\ref{App:Magnons}.  Moreover, we define $\mc{K}_s =   -i S \sigma_z \dot{R}_s^i \Gamma_i$, assuming that repeated indices, $i,j=x,y$, are summed over and we also introduce the abbreviation $\Gamma_i= \mathds{1} \pt_i  - \sigma_x \cot \Theta_0 \pt_i  \Theta_0$. The circular multiplication sign in Eq.~\eqref{ActionQ} implies convolution of the form 
\begin{equation}
\zeta^{\dagger}\circ G^{-1} \zeta\ \equiv \int_{t,\mb{r}} \int_{t',\mb{r}'} \, \zeta^{\dagger} (\mb{r},t)G^{-1}(\mb{\mb{r}}, \mb{r}',t,t') \zeta (\mb{r}',t')\,.
\end{equation}

Note that Eq.~\ref{ActionQ} assumes the absence of potentials that break translational symmetry which will generate additional classical dissipation terms \cite{Psaroudaki17} with interesting consequences on the skyrmion dynamics in confined geometries \cite{Psaroudaki18}. A considerable simplification is also provided in the limit where the skyrmion configuration energy $\mc{S}_0$ is much larger than the energy $\mc{S}_B=d \int_{\mb{r},t} \mb{b}(t) \cdot \mb{m}(\mb{r},t)$ added by the external applied field, $\mc{S}_0 \gg \mc{S}_B$. In this case, $\mb{m}(\Phi_0,\Pi_0)$ is a good approximation for the skyrmion configuration, while terms linear in the fluctuations are negligibly small and do not appear in Eq.~\eqref{ActionQ}. 

To proceed we note that the functional $\tilde{Z}$ is an integral with a Gaussian form if we neglect terms $\mc{O}[1]$ in $N_A$ originating from the Jacobian determinant $\det(J_{\mb{F}})$. Thus, after integration,  $\tilde{Z}$ reduces to 
\begin{align}
\tilde{Z}=\frac{1}{\det'( -i N_A d \hat{G}^{-1})} =\frac{e^{-\Tr' \log[1+G_0 (\tilde{\mc{K}}-\tilde{V})]}}{ \det'(-i N_A d G_0^{-1}) } \,,
\end{align}
 with $\tilde{\mc{K}}=\frac{1}{\sqrt{2}} \mc{K}_c \sigma_x + \frac{1}{\sqrt{2}} \mc{K}_q \mathds{1}$, $G_0^{-1} = \mc{G}_0^{-1} \sigma_x$, $\tilde{V} = V\sigma_x$, and the prime notation on the determinant and the trace excludes the zero modes. By performing an expansion retaining terms up to the second order in $\dot{\mb{R}}$ and first one in $V$, the effective action for the classical and quantum coordinate is 
\begin{align}
\Sef= \Scl-\frac{i}{2} \Tr'[ G_0\tilde{\mc{K}} G_0\tilde{\mc{K}} - \Delta G_0\tilde{\mc{K}} G_0\tilde{\mc{K}} - G_0\tilde{\mc{K}} \Delta G_0\tilde{\mc{K}} ]  \,,
\label{ActionEff}
\end{align}
where $\Delta G_0 = G_0 \tilde{V} G_0$. The advantage of the Keldysh rotation is that the operator $G_0$ is identified with the Green function of the fluctuations
\begin{align}
G_0= \begin{pmatrix}
G_0^K&G_0^R\\
G_0^A&0
\end{pmatrix},
\end{align}
where $G_0^{R,A}=(i S \sigma_z \partial_t \pm i0 - \mc{H})^{-1}$ are the retarded and advanced Green functions given in real time as
\begin{equation}
G_0^{R,A} (t,t') =\mp \frac{i}{S} \sigma_z \Theta(\pm(t-t')) T_{\pm} e^{-i \sigma_z \mc{H} (t-t')/S} \,,
\label{GreenFun}
\end{equation}
provided that $T_{\pm}$ time orders in chronological/antichronological order. We parametrize the Keldysh Green function as $G_0^{K} = G_0^{R} \circ F-F\circ G_0^A$, where $F=F(t-t')$ and in thermal equilibrium is given by $F(\omega) = \coth(\beta \omega/2)$, with $\beta = 1/T$. The represenation in frequency space $\omega$ is obtained by the usual Fourier transformation $g(t) =(1/2 \pi) \int_{-\infty}^{\infty}d\omega g(\omega) e^{-i \omega t}$. 

The standard way to calculate the quasiclassical equation of motion for the skyrmion coordinate $\mb{R}_c$ is to calculate the saddle point of the action \eqref{ActionEff} by extremizing with respect to the quantum coordinate $\mb{R}_q$ \cite{AltlandBook}. We note that terms proportional to $\mc{K}_q \mc{K}_c$ describe temperature-dependent dissipation due to magnon modes, while we show explicitly that terms proportional to $\mc{K}_q \mc{K}_q$ give rise to random forces. To distinguish between the contributions from these terms we  rewrite the effective action of Eq.~\eqref{ActionEff} as $\Sef= \Scl + \Sdis + \Sst$, where the dissipative part reads
\begin{align}
\Sdis= -\frac{i}{4} \Tr'& [G^{K} \mc{K}_c G^{A} \mc{K}_q + G^{R} \mc{K}_c G^{K} \mc{K}_q    \nonumber \\
&+G^{K} \mc{K}_q G^{R} \mc{K}_c + G^{A} \mc{K}_q G^{K} \mc{K}_c  ]\,,
\end{align}
where $G^{i}= G_0^{i} - \Delta G^{i}$ with $i=R,A,K$, $\Delta G ^{R,A}= G_0^{R,A} V G_0^{R,A}$ and $\Delta G^K = G^R_0 V G_0^K + G_0^K V G^A_0 $. Similarly, the stochastic part is given by
\begin{align}
\Sst & = -\frac{i}{4}  \Tr' [ G^{K} \mc{K}_q G^{K} \mc{K}_q + G^{R} \mc{K}_q G^{A} \mc{K}_q +G^{A} \mc{K}_q G^{R} \mc{K}_q ] \nonumber \\
&  \equiv -  R^{i}_q \circ C_{ij} R^{j}_q\,. 
\label{ActionSt}
\end{align}

The function $C_{ij}(t,t')$ is found by evaluating the trace appearing in Eq.~\eqref{ActionSt} with the eigenstates $\Psi_{\nu}(\mb{r},t)$ of the operator $\mc{G}$ and is given explicitly in Appendix~\ref{App:AutoCorr}. To demonstrate that $\Sst$ indeed gives rise to random fluctuating forces, we introduce auxiliary fields $\xi_{i}$ via a Hubbard-Stratonovich transformation, 
\begin{align}
e^{i \Sst}  = \det[(2 iC)^{-1}]\int \mc{D} \xi \mc{D} \xi^{\dagger} e^{i (\xi^{\dagger} \cdot (2 C)^{-1} \xi + \xi^{\dagger} \cdot \bar{R}_q  + \bar{R}_q^{\dagger} \cdot \xi)} \,,
\label{Part}
\end{align}
where $\bar{R}_q=\binom{R_q^x}{R_q^y}/\sqrt{2}$ and $\xi= \binom{\xi_x}{\xi_y}/\sqrt{2}$. Minimizing the r.h.s. of Eq.\eqref{Part} with respect to $R_q^{j}$ results in a random force term $\xi_j$ in the equation of motion characterized by an ensemble average of the form 
\begin{align}
\langle \xi(t) \rcl = 0  \,, \qquad \langle \xi_i(t) \xi_j(t') \rcl = - i C_{ij}(t,t') \,,
\label{Averages}
\end{align}
where $\langle \dots \rangle = \det[(2 iC)^{-1}]\int \mc{D} \xi \mc{D} \xi^{\dagger} \dots e^{i \xi^{\dagger} \cdot (2 C)^{-1} \xi}$. 
By minimizing the effective action $\Sef$, we obtain the dynamical Langevin equation for the classical coordinate $\mb{R}_{c}$,
\begin{align}
\tilde{Q}_0 \epsilon_{ij} \dot{R}_c^{j}(t) +\int_{-\infty}^{t} dt' ~\dot{R}_c^j (t') \gamma_{ji}(t,t') = \xi_i(t) \,,
\label{EquationTime}
\end{align}
with $\tilde{Q}_0 = -4 \pi N_A Q S d$, $\epsilon_{ij}$ is the Levi-Civita tensor and the time of preparation of the initial state is at $t \rightarrow -\infty$. The first term in Eq.~\eqref{EquationTime} is a Magnus force acting on the skyrmion and being proportional to the winding number \cite{Thiele,Stone96}, while the nonlocal (in time)  damping kernel is given by
\begin{equation}
\gamma_{ji}(t,t') =\pt_t[ \gamma_{ji}^{RK}(t,t')+\gamma_{ji}^{KA}(t,t') +\gamma_{ij}^{KR}(t',t) +\gamma_{ij}^{AK}(t',t)]\,,
\label{FrictionKernel}
\end{equation}
where 
\begin{align}
\gamma_{ji}^{ab}(t,t') =&\frac{-i S^2}{4} \sum_{\nu}\!^{}{'}\,\,\int_{\bar{\mb{r}},\mb{r},\mb{r}'} \Psi_{\nu}(\bar{\mb{r}}) G^{a}(\bar{\mb{r}},\mb{r},t,t') \sigma_z \Gamma_j(\mb{r}) \nonumber \\ & \times G^{b} (\mb{r},\mb{r}',t',t)\sigma_z \Gamma_i(\mb{r}')\sigma_z\Psi_{\nu}(\mb{r}') \,,
\end{align}
with $a,b = R,A,K$. 

The damping kernel of Eq.~\eqref{FrictionKernel} describes the dissipation which originates from the coupling of the skyrmion to the quantum bath of magnetic excitations and has an explicit temperature dependence through the Keldysh Green function $G^K$. %Finally, for later convenience we present here the Langevin equation of motion in frequency space
%\begin{align}
%\tilde{Q}_0 \epsilon_{ij}(- i \omega) R_c^{j}(\omega)  + \int_{-\infty}^{\infty} \frac{d \omega'}{2 \pi} (-i \omega') R_c^{j}(\omega') \gamma_{ji}(\omega,-\omega') =\xi_i(\omega) \,.
%\label{EquationFrequency}
%\end{align}
Note that an external force acting on the skyrmion is absent, as a direct consequence of the spatial uniformity assumed for the external magnetic field. The translational motion of the skyrmion would be induced by a spatially dependent magnetic field, for example a magnetic field gradient \cite{Wang17,Komineas15}, and its effect has been studied in Ref.~\onlinecite{Psaroudaki18}. Here, the external time-periodic field acts on the quantum bath of magnons and is naturally incorporated in the stochastic Langevin equation of Eq.~\eqref{EquationTime}. This allow us to generalize the quantum theory of dissipation to account for the effects of the driven bath in several observables related to the skyrmion dynamics.
%%%%%%%%%%%%%%%%%%%%%%%%%%%%%%%%%%%%%%%%%%%%%%%%%%%%%%%%%%%%%%%%%%%%%%%%
%%                           Quantum Dissipation
%%
%%%%%%%%%%%%%%%%%%%%%%%%%%%%%%%%%%%%%%%%%%%%%%%%%%%%%%%%%%%%%%%%%%%%%%%%  
\section{Damping Kernel}\label{sec:Dissipation} 
Our next task is to analyze the damping kernel of Eq.~\eqref{FrictionKernel} in the case of a driven bath. In Appendix \ref{App:EqDiss} we obtain the real-time damping kernel $\gamma_{ij}^0(t-t')$ in the absence of a drive, and thus establish agreement with earlier results derived in Matsubara space using the imaginary-time functional integral approach \cite{Psaroudaki17}. Note that, although the Laplace transform $\gamma^0_{ij}(z)$ is frequency dependent, we are usually interested in the long-time asymptotic behavior of the skyrmion dynamics which is in turn determined by the low frequency part of the kernel. This low frequency regime is specified by the condition $\vert \omega \vert \ll \egap$, with $\omega=\Re(i z)$, $\egap = 2\kappa +h$, being the lowest magnon gap, while at the same time the temperature is limited to the quantum regime $T \ll \egap$. 
%%%%%%%%%%%%%%%%%%%%%%%%%%%%%%%%%%%%%%%%%%%%%%%%%%%
\begin{figure}[t]
\centering
\includegraphics[width=1\linewidth]{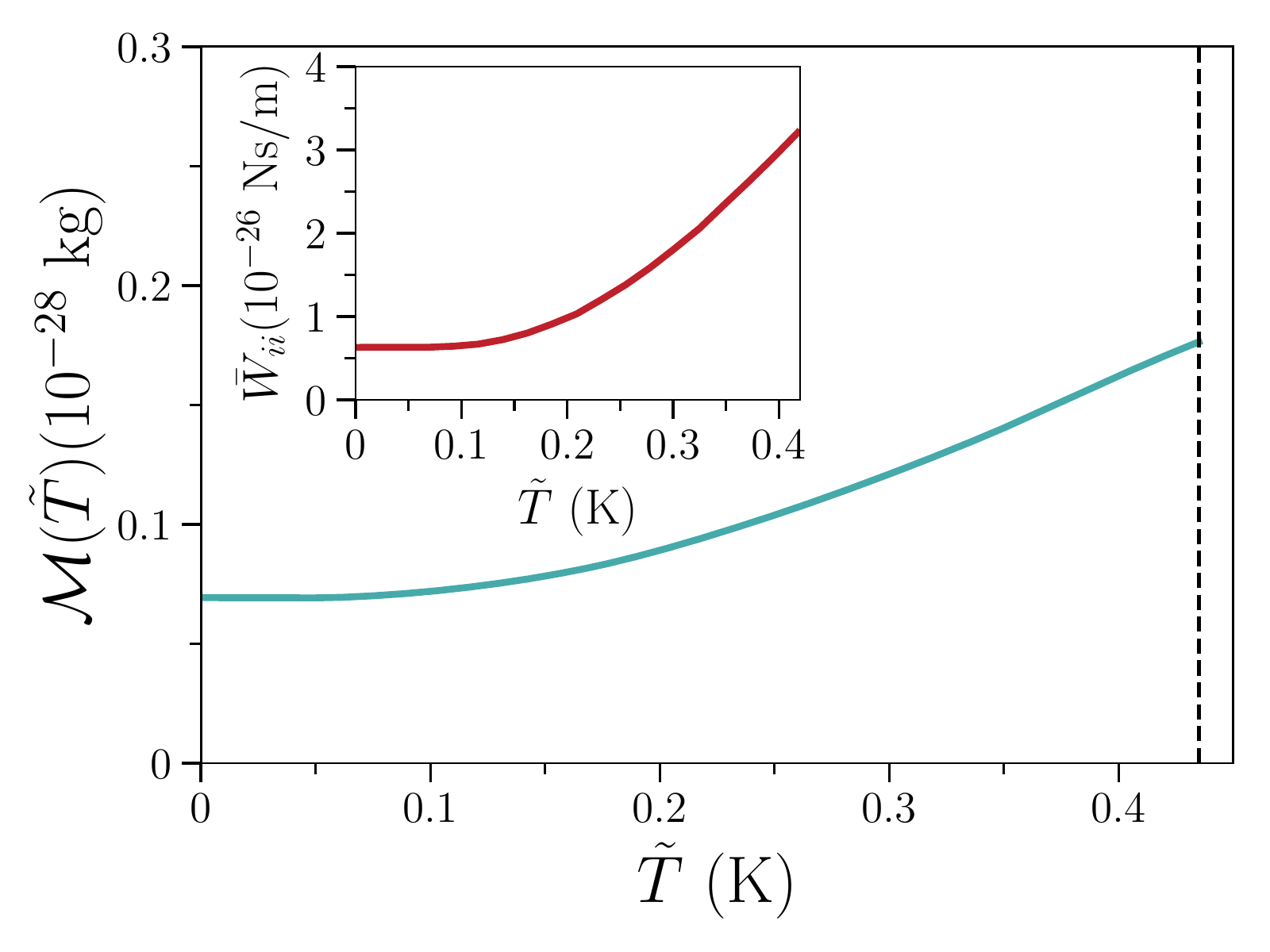}
\caption{Temperature dependence of the quantum mass $\mc{M}(\tilde{T})$ of the skyrmion given in Eq.~\eqref{MassT} for a static magnetic field of amplitude $H=216$ mT, radius $\lambda=5.32$ nm, and a choice of $J=1$ meV, $S=1$ and $J/D=4$, and $\alpha=5$ \AA. The dashed vertical line indicates the value of the magnon gap in units of temperature $\egap=0.435$ K, up to which our result is valid. The inset depicts the constant $\bar{W}_{ii}$ of Eq.~\eqref{W0} for an oscillating magnetic field of amplitude $b_0 = 0.05$ (27 mT), $\wext=0.32$ (4.8 GHz), and $\phiext=\pi/4$. 
} \label{Mass0}
\end{figure}
%%%%%%%%%%%%%%%%%%%%%%%%%%%%%%%%%%%%%%%%%%%%%%%%%%%

Thus, under the assumptions specified above the diagonal damping kernel acquires the super-Ohmic power law behavior $\gamma^0_{ii}(z) =z \mc{M}(T)  + \mc{O}[(z/\egap)^2]$. Following the usual terminology \cite{WeissBook}, Ohmic friction is described by a damping term of the form $z \gamma (z) \propto z^s$ with $s=1$, while for $s>1$ we call it super-Ohmic. The $T$-dependent mass is given by,
\begin{align}
\mc{M}(T)= \sum_{\nu,\nu'}\!^{}{'}\,\, \frac{\Re(\mc{B}_{ii}^{\nu\nu'}) \bar{F}_{\nu\nu'}}{\varepsilon_{\nu'}-\varepsilon_{\nu}} \,,
\label{MassT}
\end{align}
with $\bar{F}_{\nu\nu'} = F(\varepsilon_{\nu})- F(\varepsilon_{\nu'})$ and $F(\varepsilon_{\nu}) =\coth(\beta \varepsilon_{\nu}/2)$. Here, the sum runs over the quantum number $\nu = \{q=\pm 1,n\}$, where the index $q$ distinguishes between particle states ($q=1$), solutions of the eigenvalue problem $\mc{H} \Psi_n = \varepsilon^q_n \sigma_z \Psi_n$ with positive eigenfrequency $\varepsilon_n^{1} =+ \varepsilon_n$, and antiparticle states ($q=-1$) with negative eigenfrequency $\varepsilon_n^{-1} =- \varepsilon_n$ \cite{Psaroudaki17}. The matrix elements are given by $\mc{B}_{ij}^{\nu\nu'} = \mc{B}_{ij}^{n,q;n',q'} =  (q q'/2)  \int_{\mb{r}} \Psi_{\nu}^{\dagger} \Gamma_i \sigma_z\Psi_{\nu'} \int_{\mb{r'}} \Psi_{\nu'}^{\dagger} \Gamma_j \sigma_z \Psi_{\nu}$. Note that the expression of Eq.~\eqref{MassT} is symmetric under the exchange of indices $\nu$ and $\nu'$, and that there is no singularity for $\varepsilon_\nu=\varepsilon_{\nu'}$ since $\lim_{\varepsilon_\nu \rightarrow \varepsilon_{\nu'}} \bar{F}_{\nu' \nu}/(\varepsilon_\nu - \varepsilon_{\nu'})=\beta/2\sinh^2(\beta \varepsilon_\nu/2)$. The quantum nature of the magnon bath is evident from the non-vanishing $\mc{M}(T)$ in the $T \rightarrow 0$ limit. In order to emphasize that $\mc{M}(0)$ is finite and that it is independent of the effective spin $N_A S$, contrary to the magnus force proportional to $\tilde{Q}_0= -4 \pi Q N_A S d$, we refer to the mass of Eq.~\eqref{MassT} as \textit{quantum mass}. This terminology allows us to distinguish $\mc{M}(T)$ from the semiclassical mass already calculated in Ref.~\onlinecite{Psaroudaki17} in the presence of spatial confinement, which scales linearly with $N_A S$.

The off-diagonal damping kernel has a super-Ohmic low-frequency power law $\gamma_{xy}(z) \propto z^2$, irrelevant for the skyrmion dynamics at times $t \gg \egap^{-1}$. The $T$-dependence of the quantum mass $\mc{M}(T)$ is depicted in Fig.~\ref{Mass0} in physical units for $J=1$ meV, $\alpha=5$ \AA, $J/D=4$, $h=0.4$ (216 mT), $\lambda=2.67$ (5.34 nm), $\kappa=0.1$, and $S=1$. Details on the calculation are given in Appendix~\ref{App:Magnons}.

With this preparation, we are now in position to generalize the damping kernel in the presence of the external driving field turned on at time $t=t_0$, $\mb{b}(t)= b_0 \Theta(t-t_0) \cos(\wext t ) ( \sin \phiext,0, \cos \phiext)$, tilted in the $xz$-plane with the angle $\phiext$ away from the $z$-axis. In the presence of $\mb{b}(t)$, the magnons are subjected to the potential $V(\mb{r},t) = b_0 \Theta(t-t_0) \cos(\wext t) V(\mb{r})$, where $V(\mb{r})$ is given in Eq.~\eqref{PotentialDrive}. The damping kernel of Eq.~\eqref{FrictionKernel} acquires an additional correction due to the time-dependent field, $\gamma_{ji}(t,t') = \gamma_{ji}^{0}(t-t') + \Delta \gamma_{ji}(t, t')$, where 
\begin{equation}
\Delta \gamma_{ji}(t, t') =  \partial_{t} W_{ji}(t-t')[\gex(t)+ \gex(t')]\,.
\label{DampingCor}
\end{equation}
The function $\gex(t)$ carries information on the external drive, $\gex(t)= \Theta(t-t_0) b_0 \cos(\wext t - \vert \epsilon_{ji} \vert \pi/2)$, while $W_{ji}(t)$ carries information about the magnon modes,
\begin{equation}
W_{ji}(t) ={\sum_{\nu_1, \nu_2,\nu_3}}\!\!\!\!^{}{'}\,\,\frac{\mc{C}_{ji}^{\nu_1\nu_2\nu_3}[ w_{\nu_3\nu_2}(t)-w_{\nu_3\nu_1}(t)]}{(\varepsilon_{\nu_2}- \varepsilon_{\nu_1})^2 -\wext^2}  \,,
\label{DampingTimeCor}
\end{equation}
where $w_{\nu_1\nu_2}(t) = \Theta(t) \bar{F}_{\nu_1\nu_2} \sin[(\varepsilon_{\nu_1}-\varepsilon_{\nu_2}) t]$. We also introduced the matrix elements
\begin{align}
\mc{C}_{ii}^{\nu_1\nu_2\nu_3} &= q_{\nu_1} q_{\nu_2} q_{\nu_3} (\varepsilon_{\nu_2}-\varepsilon_{\nu_1}) \Re(b_i^{\nu_3\nu_1} V_{\nu_1\nu_2} b_i^{\nu_2\nu_3})/2S\,, \nonumber \\
\mc{C}_{yx}^{\nu_1\nu_2\nu_3} &= q_{\nu_1} q_{\nu_2} q_{\nu_3} \wext \Im(b_y^{\nu_3\nu_1}V_{\nu_1\nu_2} b_x^{\nu_2\nu_3})/2S \,,
\label{MatrixEl}
\end{align}
where $b_i^{\nu_1\nu_2} = \int_{\mb{r}} \Psi^{\dagger}_{\nu_1} \Gamma_i \sigma_z \Psi_{\nu_2}$ and $V_{\nu_1\nu_2} = \int_{\mb{r}}  \Psi^{\dagger}_{\nu_1} V \Psi_{\nu_2}$. We note that the triple summation over the magnon quantum numbers originates from the fact that the external field induces a finite overlap, $V_{\nu_1\nu_2}\neq 0$ for $\nu_1 \neq \nu_2$. Note that Eq.~\eqref{DampingTimeCor} is valid only away from the resonance condition $\wext = \varepsilon_{\nu_2}-\varepsilon_{\nu_1}$, under the assumption that the external potential $V$ induces only a small overlap $0< \vert V_{\nu_1\nu_2}\vert \ll 1$ between magnon modes carrying approximately the same energy. Thus, the energy differences are restricted as $0 \leq \vert \varepsilon_{\nu_1}- \varepsilon_{\nu_2} \vert \leq \varepsilon_d $ and it also holds that $\varepsilon_d \ll \wext$.

In Fourier  space with frequency $\omega$, the equation of motion given in Eq.~\eqref{EquationTime}  takes the form
\begin{align}
F(t,\omega) = -i\omega [\tilde{Q}_0 \epsilon_{ij} +\gamma_{ji}(t,\omega)]R_c^{j}(\omega) - \xi_i(\omega) \,,
\label{EquationFrequency}
\end{align}
with $F(t,\omega)$ satisfying $\int_{-\infty}^{\infty} d \omega e^{-i \omega t} F(t,\omega) = 0$,  and where $\gamma_{ji}(t,\omega) = \gamma_{ji}^0(\omega) + \Delta \gamma_{ji}(t,\omega)$. It appears convenient to calculate $\Delta \gamma_{ji}(t,\omega)$ in Laplace space $z$ with $\omega=\Re(i z)$, 
\begin{align} 
~&\Delta \gamma_{ji}(t,z) = W_{ji}(z) [z \gex(t) +\pt_t\gex(t)]   \label{NonEqDissi} \\ 
 &+ \frac{b_0}{2}\sum_{m=\pm 1} e^{-i m (\wext t+\frac{\pi \vert \epsilon_{ji}\vert}{2}) }(z +i m \wext)W_{ji}(z+i m \wext)  \nonumber \,.
\end{align}
The correction to the damping kernel,  $\Delta \gamma_{ji}(t,z)$, describes the effects of the driven magnon bath on the skyrmion and is treated as a perturbation to $\gamma_{ji}^0(z)$. Here, $W_{ji}(z)$ is the Laplace transform of $W_{ji}(t)$ given in Eq.~\eqref{DampingTimeCor}. In Eq.~\eqref{EquationFrequency} we assume that the time $t_0$ coincides with the preparation time of the initial state, i.e. $t_0 \rightarrow -\infty$, and we therefore neglect boundary terms that depend on $t_0$. A Taylor expansion around the origin,  $\gamma_{ji}(t,z) \simeq \gamma_{ji}(t,0) +z \pt_{z}  \gamma_{ji}(t,z)\vert_{z=0} + \mc{O}(z^2)$, valid for frequencies $\omega \ll \egap$, %with $\omega = \Re(i z)$, 
provides the low frequency power-law behavior of the damping kernel. For the diagonal part we find 
\begin{equation}
\Delta \gamma_{xx}(t,z) \simeq D(T)\sin(\wext t) +z~\delta M(T)\cos(\wext t)\,,
\label{Diagonal}
\end{equation}
and similarly the off-diagonal corrections are
\begin{equation}
\Delta \gamma_{yx}(t,z) \simeq \delta Q(T)\cos(\wext t) +z~ G(T)\sin(\wext t)\,.
\label{offDiagonal}
\end{equation}

Explicit expressions of the $T$-dependent coefficients appearing in Eqs.~\eqref{Diagonal} and~\eqref{offDiagonal} are given in Appendix~\ref{App:NonEqDiss}. As expected, in the static limit $\wext \rightarrow 0$, all the terms in Eqs.~\eqref{Diagonal} and \eqref{offDiagonal}, except the mass renormalization, vanish.  In the special case of $\varepsilon_{d} \ll \wext \ll \egap$, where $0\leq \vert \varepsilon_{\nu_2}- \varepsilon_{\nu_1} \vert \leq \varepsilon_{d}$ is the energy difference induced by the external potential $V$, we find the simplified expressions $D(T) = -\wext \bar{W}_{ii}$, $\delta M(T) = \bar{W}_{ii}$, $\delta Q(T) = \wext \bar{W}_{yx}$, and $G(T) = \bar{W}_{yx}$. The  coefficient $\bar{W}_{ji}$ is given by
\begin{equation}
\bar{W}_{ji}= \sum_{\nu_1, \nu_2,\nu_2}\!\!\!\!\!^{}{'}\,\,\frac{2 \mc{C}_{ji}^{\nu_1\nu_2\nu_3}}{\wext^2}\left( \frac{\bar{F}_{\nu_2\nu_3}}{\varepsilon_{\nu_3}- \varepsilon_{\nu_2} }-\frac{\bar{F}_{\nu_1\nu_3}}{\varepsilon_{\nu_3}- \varepsilon_{\nu_1} }\right) \,,
\label{W0}
\end{equation} 
where $\bar{F}_{\nu\nu'}$ is given after Eq.~\eqref{MassT}. From Eq.~\eqref{W0} and the structure of the matrix elements of Eq.~\eqref{MatrixEl} it becomes apparent that $\bar{W}_{ji}$ is symmetric under the exchange of the indices $\nu_1$, $\nu_2$, and $\nu_3$. The temperature dependence of the coefficient $\bar{W}_{ii}$ is depicted in the inset of Fig.~\ref{Mass0}, for the choice $\phiext=\pi/4$, $b_0 =0.05$ (27 mT), $\wext =0.32$ (4.8 GHz), and $h=0.4$ (216 mT). 

Due to the symmetries of the matrix elements we note the relations $\Delta \gamma_{xx}(t,z) = \Delta \gamma_{yy}(t,z)$ and $\Delta \gamma_{xy}(t,z) = - \Delta \gamma_{yx}(t,z)$, thus the term $\delta Q (T) \cos(\wext t)$ can be considered as a temperature- and time-dependent correction to the topological charge $\tilde{Q}_0$, induced by the external drive. Similarly, the quantum mass acquires the correction $\delta M(T) \cos(\wext t)$. The low-frequency linear dependence of the quantity $ z \gamma_{ji}(t,z)$ signals a super-Ohmic to Ohmic crossover behavior, with measurable consequences on the skyrmion trajectory~\cite{Psaroudaki18}. More specifically, the ac driving of the magnon bath at resonance displaces the skyrmion from its equilibrium position and results in a unidirectional helical propagation.
%%%%%%%%%%%%%%%%%%%%%%%%%%%%%%%%%%%%%%%%%%%%%%%%%%%%%%%%%%%%%%%%%%%%%%%%
%%                           Section : Response Function 
%%
%%%%%%%%%%%%%%%%%%%%%%%%%%%%%%%%%%%%%%%%%%%%%%%%%%%%%%%%%%%%%%%%%%%%%%%%
%%%%%%%%%%%%%%%%%%%%%%%%%%%%%%%%%%%%%%%%%%%%%%%%%%%
\begin{figure}[t]
\centering
\includegraphics[width=1\linewidth]{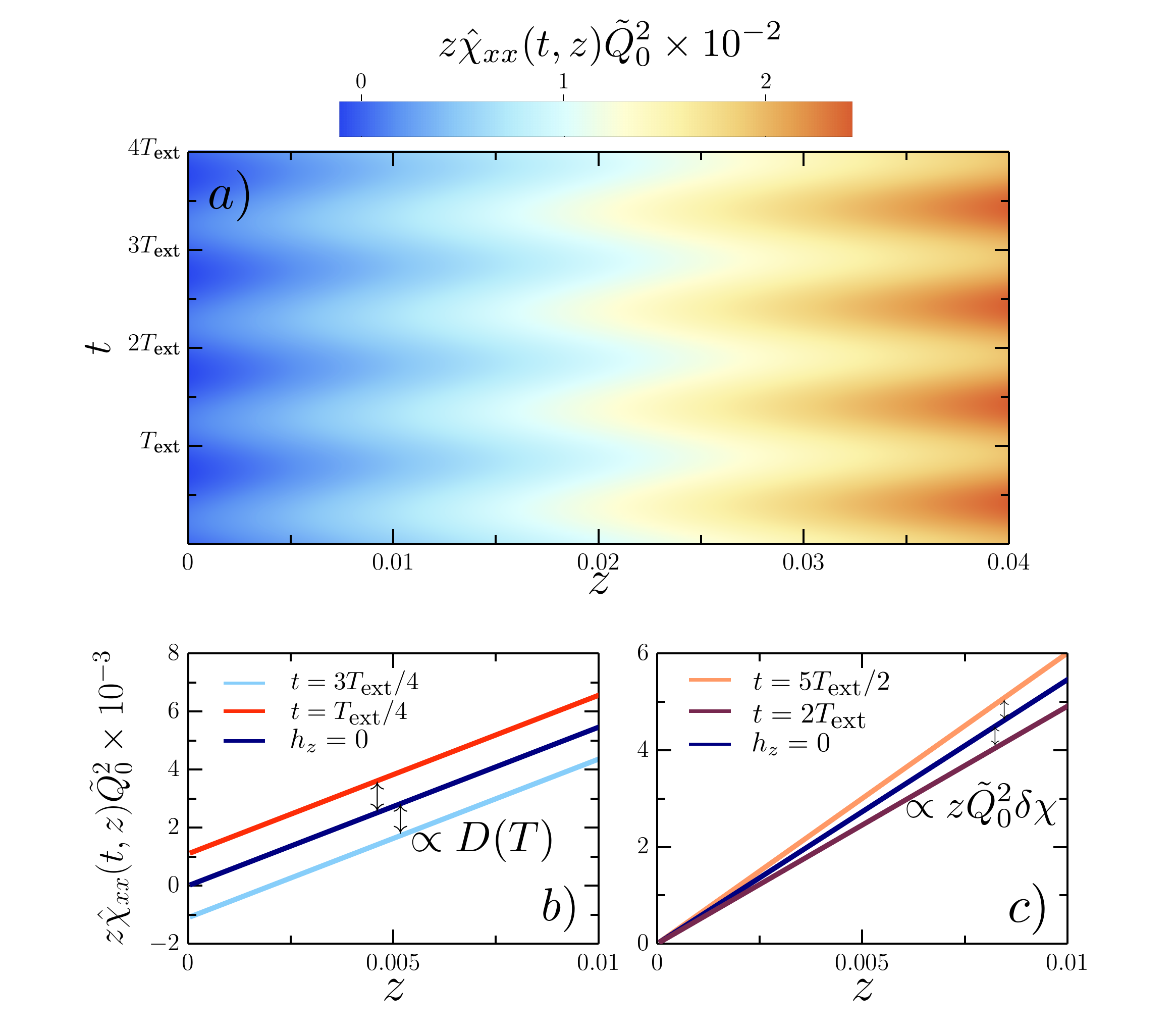}
\caption{$a)$ The colored surface represents the time and Laplace frequency dependence of the diagonal response function $\chi_{xx}(t,z)$, given in Eq.~\eqref{ResponseDiag}, for $T=0.4$ (0.3 K), $\wext =0.32$ (4.8 GHz), $\phiext=\pi/4$, and $b_0 =0.05$ (27 mT) the amplitude of the external field. The skyrmion is stabilized from a uniform out-of-plane magnetic field of strength $h=0.4$ (216 mT) and has a radius of $\lambda=2.67$ (5.32 nm), and we choose $J=1$ meV, $\alpha=0.5$ nm and $J/D=4$. Insets $b)$ and $c)$ depict the frequency dependence of  $\chi_{xx}(t,z)$ at given times $t$, where $T_{\mbox{\tiny{ext}}}=2 \pi/ \wext$ denotes the period of the external drive. 
} \label{chi_xx}
\end{figure}
%%%%%%%%%%%%%%%%%%%%%%%%%%%%%%%%%%%%%%%%%%%%%%%%%%%
%%%%%%%%%%%%%%%%%%%%%%%%%%%%%%%%%%%%%%%%%%%%%%%%%%%
\begin{figure}[t]
\centering
\includegraphics[width=1\linewidth]{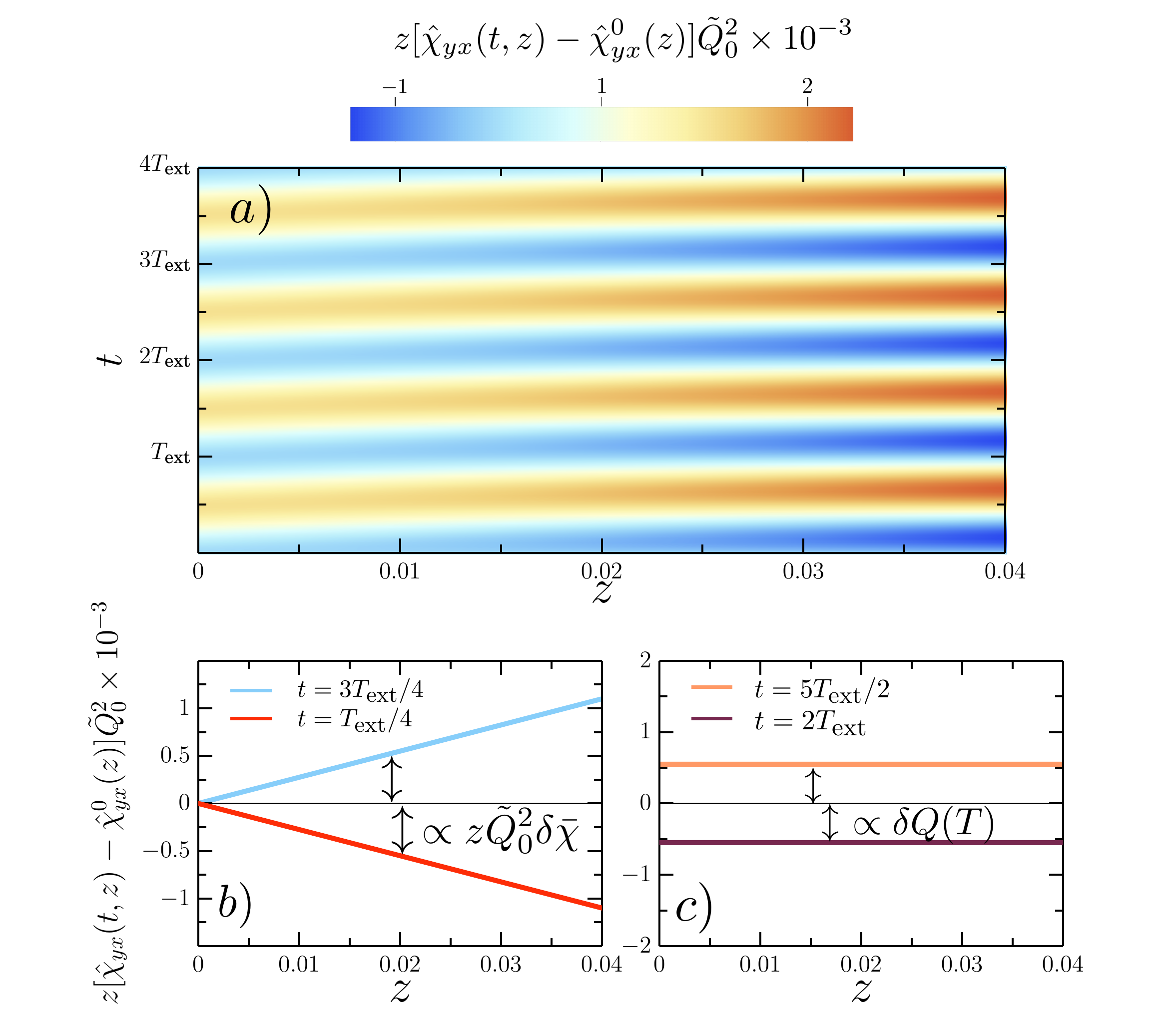}
\caption{$a)$ The colored surface represents the time and Laplace frequency dependence of the off-diagonal response function $\chi_{yx}(t,z)$, given in Eq.~\eqref{ResponseoffDiag} for $T=0.4$ (0.3 K), $\wext =0.32$ (4.8 GHz), $\phiext=\pi/4$ and $b_0 =0.05$ (27 mT) the amplitude of the external field. The skyrmion is stabilized from a uniform out-of-plane magnetic field of strength $h=0.4$ (216 mT) and has a radius of $\lambda=2.67$ (5.32 nm), and we choose $J=1$ meV, $\alpha=0.5$ nm and $J/D=4$. Insets $b)$ and $c)$ depict the frequency dependence of  $\chi_{xx}(t,z)$ at given times $t$, where $T_{\mbox{\tiny{ext}}}= 2 \pi/\wext $ denotes the period of the external drive. 
} \label{chi_yx}
\end{figure}
%%%%%%%%%%%%%%%%%%%%%%%%%%%%%%%%%%%%%%%%%%%%%%%%%%%
\section{Response Function}\label{sec:Response} 
In this section, we calculate the equilibrium skyrmion response function, which is then generalized to the nonequilibrium case of a driven bath of magnons. The linear response of the skyrmion to the fluctuating force $\xi_i(t)$ is encoded in the equilibrium response function $\chi^{0}_{ij}(t-t')$ via the relation 
\begin{align}
R^i_c(t) = \int_{-\infty}^{t} dt' \chi^{0}_{ij}(t-t') \xi_j(t') \,,
\label{ResponseDef}
\end{align}
where the elements in Laplace space are
\begin{align}
\chi^0_{ii}(z) = \frac{\gamma^0_{ii}(z)}{z \pi^{0}(z)} \,,\quad \chi^0_{yx}(z) =\frac{\tilde{Q}_0+\gamma^0_{yx}(z)}{z \pi^{0}(z)} \,,
\label{Response}
\end{align}
with $\pi^{0}(z)= [\tilde{Q}_0+\gamma_{yx}^0(z)]^2+[\gamma_{xx}^0(z)]^2$ and $\chi^0_{xy}(z) = - \chi^0_{yx}(z)$. The response functions at finite frequency $\chi^0_{ij}(z)$ are dynamical observables carrying physical information on the skyrmion dynamics. By employing the low-frequency power law behavior of $\gamma^0_{ij}(z)$, one finds that $\chi^0_{ii}(z)= \mc{M}(T)/[\tilde{Q}_0^2+\mc{M}(T)^2 z^2]$ and $\chi^0_{yx}(z)= \tilde{Q}_0/[z(\tilde{Q}_0^2+\mc{M}(T)^2 z^2)]$. The expansion at $z=0$ yields $\chi^0_{ii}(z) \simeq \mc{M}(T)/\tilde{Q}_0^2 +\mc{O}(z^2)$. A finite static susceptibility $\chi_0 = \mc{M}(T)/\tilde{Q}_0^2$ implies that a free topological particle with $\tilde{Q} \neq 0$ exhibits a different dynamical behavior than the one with $\tilde{Q}_0 =0$. In particular, we note that the static susceptibility $\chi_0$ is infinite for a freely moving and finite for a confined Brownian particle \cite{WeissBook}. For example, $\chi_0 = 1/\omega_0^2$ for a damped harmonic oscillator of frequency $\omega_0$ \cite{WeissBook}. Therefore, we see that $\chi_0$ is finite due to the non-trivial $\tilde{Q}_0$, and as expected, $\chi_0$ diverges for $\tilde{Q}_0 = 0$. Moreover, the low frequency expansion for the off-diagonal response function is $\chi^0_{yx}(z) \simeq (1/\tilde{Q}_0 z) +\mc{O}(z)$, and in this case $\tilde{Q}_0$ plays the role of a velocity-dependent friction.

The response of the skyrmion position ${\bf R}(t)$ when the external drive $\mb{b}(t)$ is turned on is encoded in the response function $\chi_{ij}(t,t')$ defined through the relation
\begin{align}
R^i_c(t) = \int_{-\infty}^{t} dt' \chi_{ij}(t,t') \xi_j(t') \,. 
\label{ResponseGen}
\end{align} 
In an analogous fashion to the decomposition of the damping kernel given in Eq.~\eqref{DampingCor}, we generalize the response function as $ \chi_{ij}(t,t') =  \chi_{ij}^{0}(t-t') + \delta  \chi_{ij}(t,t')$. Starting from the equation of motion given in Eq.~\eqref{EquationTime} and using Eq.~\eqref{ResponseGen}, we solve for the function $\delta \chi_{ij}(t,\omega)$, defined as $\delta \chi_{ij}(t,t') = (1/2\pi) \int d\omega e^{-i\omega(t-t')}\delta \chi_{ij}(t,\omega)$, retaining first order terms in $b_0$. In Laplace space, by performing an expansion of the full response function $ \chi_{ij}(t,z) = \chi^0_{ij}(z) + \delta \chi_{ij}(t,z)$ around $z=0$ and keeping leading order terms in $z$, we find
\begin{equation}
\chi_{ii} (t,z) \simeq \frac{D(T)}{\tilde{Q}_0^2 z} \sin(\wext t) +\chi_0 +\delta \chi \cos(\wext t) \,,
\label{ResponseDiag}
\end{equation}
with $\delta\chi=[-2 \mc{M}(T) \delta Q(T) +\tilde{Q}_0 \delta M(T)]/\tilde{Q}_0^3$, and similarly 
\begin{equation}
\chi_{yx} (t,z) \simeq \frac{1}{\tilde{Q}_0z} - \frac{\delta Q(T)}{\tilde{Q}_0^2 z}\cos(\wext t)  +\delta \bar{\chi} \sin(\wext t)\,,
\label{ResponseoffDiag}
\end{equation}
where $\delta \bar{\chi} = [-2 \mc{M}(T) D(T) -\tilde{Q}_0 G(T)]/\tilde{Q}_0^3$. We observe that a new friction term emerges for the diagonal response function and a new static susceptibility term for the off-diagonal one. The characteristic behavior of the response functions $\chi_{ji}(t,z)$ is illustrated in Figs.~\ref{chi_xx}--\ref{chi_yx}. To begin with, an anticipated result is depicted in the colored surfaces plotted in Figs.~\ref{chi_xx}(a) and \ref{chi_yx}(a), namely that $\chi_{ji}(t,z)$ are periodic functions of time $t$, with a period $ \Text = 2\pi/\wext = 19.63$ (1.3 ns). The $z$-dependence of $\chi_{ji}(t,z)$ carries information on the memory effects that originate from the skyrmion-magnon bath coupling, including the additional dissipative terms generated by the oscillating driving field. Thus we notice that the diagonal $\chi_{ii}(t,z)$ depends on the friction coefficient $D(T)$, while the off-diagonal $\chi_{yx}(t,z)$ has a dependence on the topological charge renormalization $\delta Q(T)$.

 %%%%%%%%%%%%%%%%%%%%%%%%%%%%%%%%%%%%%%%%%%%%%%%%%%%%%%%%%%%%%%%%%%%%%%%%
%%                           Fluctuation - Dissipation Theorem
%%
%%%%%%%%%%%%%%%%%%%%%%%%%%%%%%%%%%%%%%%%%%%%%%%%%%%%%%%%%%%%%%%%%%%%%%%%
\section{Fluctuation-Dissipation theorem}\label{sec:FDT} 
In this section we turn our attention to the derivation of the fluctuation-dissipation (FD) theorem, for a skyrmion in contact to a bath of magnons at equilibrium. An extension of the FD relation is also derived for a nonequilibrium bath of magnons which is weakly driven by an oscillating magnetic field, a relation which reduces to the FD theorem in the static limit. We also calculate the time and temperature dependence of the skyrmion mean square displacement (MSD).

The FD theorem relates equilibrium thermal fluctuations and dissipative transport coefficients \cite{WeissBook,Hanggi05}. In the absence of an external drive, the Fourier transform $C^0_{ij}(\omega)$ of the quantum stochastic force correlation function defined through Eq.~\eqref{ActionSt} is related to the damping kernel $\gamma^0_{ij}(\omega)$ by the relation, 
\begin{equation}
C^0_{ij}(\omega) +C^0_{ji}(-\omega)=  i\omega \coth(\frac{\beta \omega}{2}) [\gamma^0_{ij}(\omega) +\gamma^0_{ji}(-\omega)] \,.
\label{powerspec}
\end{equation}
Eq.~\eqref{powerspec} is the quantum mechanical version of the FD theorem with the observation that quantum effects enter not only through the usual $\omega \coth(\beta \omega/2)$ term, but additionally through the non-trivial $\propto \coth(\beta \varepsilon_\nu/2)$ dependence of the damping kernel $\gamma_{ij}(\omega)$. 

We now turn to the extension of the FD relation of Eq.~\eqref{powerspec} in the presence of an external field $\mb{b}(t)$. In general, the stochastic fluctuations of reservoirs driven out of equilibrium do not necessarily relate to their dissipative properties, and a generalization of the FD theorem should not be expected, except for some special cases \cite{Sukhorukov01,Altland01}. Following the same methodology as in Sec.~\ref{sec:Dissipation}, we decompose the random force autocorrelation function as follows,
\begin{align}
\langle \xi_{j}(t) \xi_{i}(t') \rangle = -i[C^{0}_{ji}(t-t') + \Delta C_{ji}(t,t')]\,,
\end{align}
where the stochastic function $\Delta C_{ji}(t,t')$ satisfies
\begin{align}
\Delta C_{ji}(t,t') = \pt_{t} \pt_{t'} U_{ji}(t-t') [\gex(t) + \gex(t')] \,.
\end{align}
Here, $U_{ji}(t-t')$ carries information about the magnon bath and is given by
\begin{align}
U_{ji}(t) = \sum_{\nu_1, \nu_2,\nu_2}\!\!\!\!^{}{'}\,\,\frac{i \mc{C}_{ji}^{\nu_1\nu_2\nu_3}[ u_{\nu_3\nu_2}(t)-u_{\nu_3\nu_1}(t)]}{2[(\varepsilon_{\nu_2}- \varepsilon_{\nu_1})^2 -\wext^2]}  \,,
\label{FluctuationTimeCor}
\end{align}
where $u_{\nu_1\nu_2}(t) = [1-F(\varepsilon_{\nu_1})F(\varepsilon_{\nu_2})] \cos[(\varepsilon_{\nu_1}-\varepsilon_{\nu_2}) t]$. We remind the reader that the damping kernel $\Delta \gamma_{ji}(t, t')$ equals $\Delta \gamma_{ji}(t,t') = \pt_{t} W_{ji}(t-t') [\gex(t) + \gex(t')]$, with $W_{ji}(t)$ given in Eq.~\eqref{DampingTimeCor}. The generalization of the FD theorem is found to be independent of the form of the external drive and is expressed as a relation between the functions $W_{ji}(t)$ and $U_{ji}(t)$ in Fourier space,
\begin{equation}
U_{ji}(\omega) + U_{ij}(-\omega) =\coth(\frac{\beta \omega}{2})[W_{ji}(\omega) - W_{ij}(-\omega)] \,. 
\label{FDRelation}
\end{equation}
The non-equilibrium FD relation Eq.~\eqref{FDRelation} is valid within first order perturbation theory with respect to the amplitude of the driving field, however we expect it will serve as a basis for future investigations of the effects of time-dependent driving fields beyond first-order perturbation theory. In the special case of a static external field $\wext \rightarrow 0$, the FD theorem in equilibrium, Eq.~\eqref{powerspec}, is recovered trivially,
\begin{equation}
C_{ij}(\omega) +C_{ji}(-\omega)= i \omega \coth(\frac{\beta \omega}{2}) [\gamma_{ij}(\omega) +\gamma_{ji}(-\omega)] \,,
\label{NonEqFDT}
\end{equation}
where $C_{ij}(\omega) = C^0_{ij}(\omega)+2\omega^2 U_{ij}(\omega)\gex(0)$ and $\gamma_{ij}(\omega) = \gamma^0_{ij}(\omega) + 2(-i \omega) W_{ij}(\omega)\gex(0)$.  
%%%%%%%%%%%%%%%%%%%%%%%%%%%%%%%%%%%%%%%%%%%%%%%%%%%
\begin{figure}[t]
\centering
\includegraphics[width=1\linewidth]{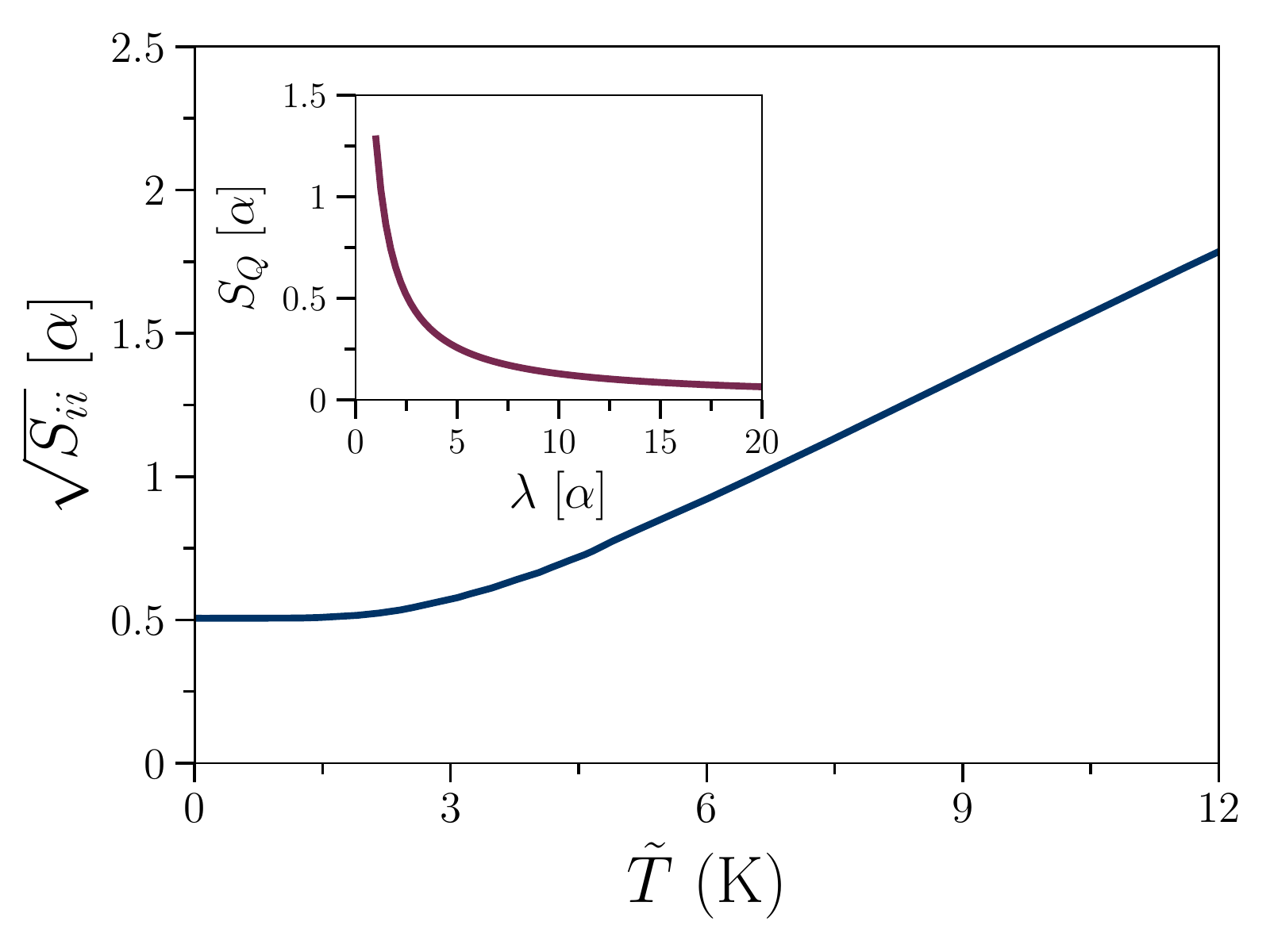}
\caption{Root mean square displacement (RMSD) $\sqrt{S_{ii}}$ given in Eq.~\eqref{MSD} as a function of temperature $\tilde{T}$ at time $t=6.6$ ps, for a skyrmion of radius $\lambda=2.57\alpha$, and $Q_0=-1$. The RMSD is plotted for the choice $J=1$ meV, $N_A S=1$, and $d =1$, and given in units of the lattice constant $\alpha$. Due to the quantum magnetic excitations, the RMSD at zero temperature, $S_Q\equiv \sqrt{S_{ii}}(\tilde{T}=0)$, remains finite, while it scales linearly with $\tilde{T}$ at finite temperatures. The inset depicts the dependence of $S_Q$ on the skyrmion size $\lambda$.} \label{sT}
\end{figure}
%%%%%%%%%%%%%%%%%%%%%%%%%%%%%%%%%%%%%%%%%%%%%%%%%%%

We now focus on  the temperature dependence of the r.h.s. of Eq.~\eqref{powerspec}, which we expect to give rise to a finite zero-temperature mean squared displacement (MSD) of the skyrmion position. This motivates us to consider the correlation function $S_{ij}(t,t')=\frac{1}{2}\langle [R_{i}(t)- R_j(t')]^2\rcl $, where $ \langle  \ldots \rangle$ denotes ensemble average, and where $ \langle  \mb{R}\rangle = 0$. From Eqs.~\eqref{Response} and \eqref{Averages} it follows that in the special case of $\mb{b}(t)=0$, the diagonal MSD $S_{ii}(\bar{t})=S_{ii}(t-t')$ reduces to
\begin{equation}
S_{ii}(\bar{t}) = \int \frac{d \omega}{2 \pi} (e^{-i \omega\bar{t}}-1)\chi_{il}(\omega) \mc{X}_{lk}(\omega) \chi_{ik}(-\omega) \,,
\label{MSD}
\end{equation} 
where $\mc{X}^0_{ij}(\omega) = -i [C^0_{ij}(\omega) +C^0_{ji}(-\omega)]$ is the symmetrized autocorrelation function. Eq.~\eqref{MSD} contains several contributions, 
%which for brevity, we omit. From these, 
of which we retain only the leading terms in $\tilde{Q}_0$, under the assumption $\tilde{Q}_0 \gg 1$, to further simplify the MSD to
\begin{equation}
S_{ii}(\bar{t})= \frac{2 \pi}{\tilde{Q}_0^2}\sum_{\nu,\nu'}\!^{}{'}\,\, \mc{B}_{ii}^{\nu;\nu'}[F(\varepsilon_{\nu})F(\varepsilon_{\nu'})-1] \sin^2[(\varepsilon_{\nu'}-\varepsilon_{\nu}) \bar{t}/2] \,. 
\label{MeanSquare}
\end{equation}
%%%%%%%%%%%%%%%%%%%%%%%%%%%%%%%%%%%%%%%%%%%%%%%%%%%
\begin{figure}[t]
\centering
\includegraphics[width=1\linewidth]{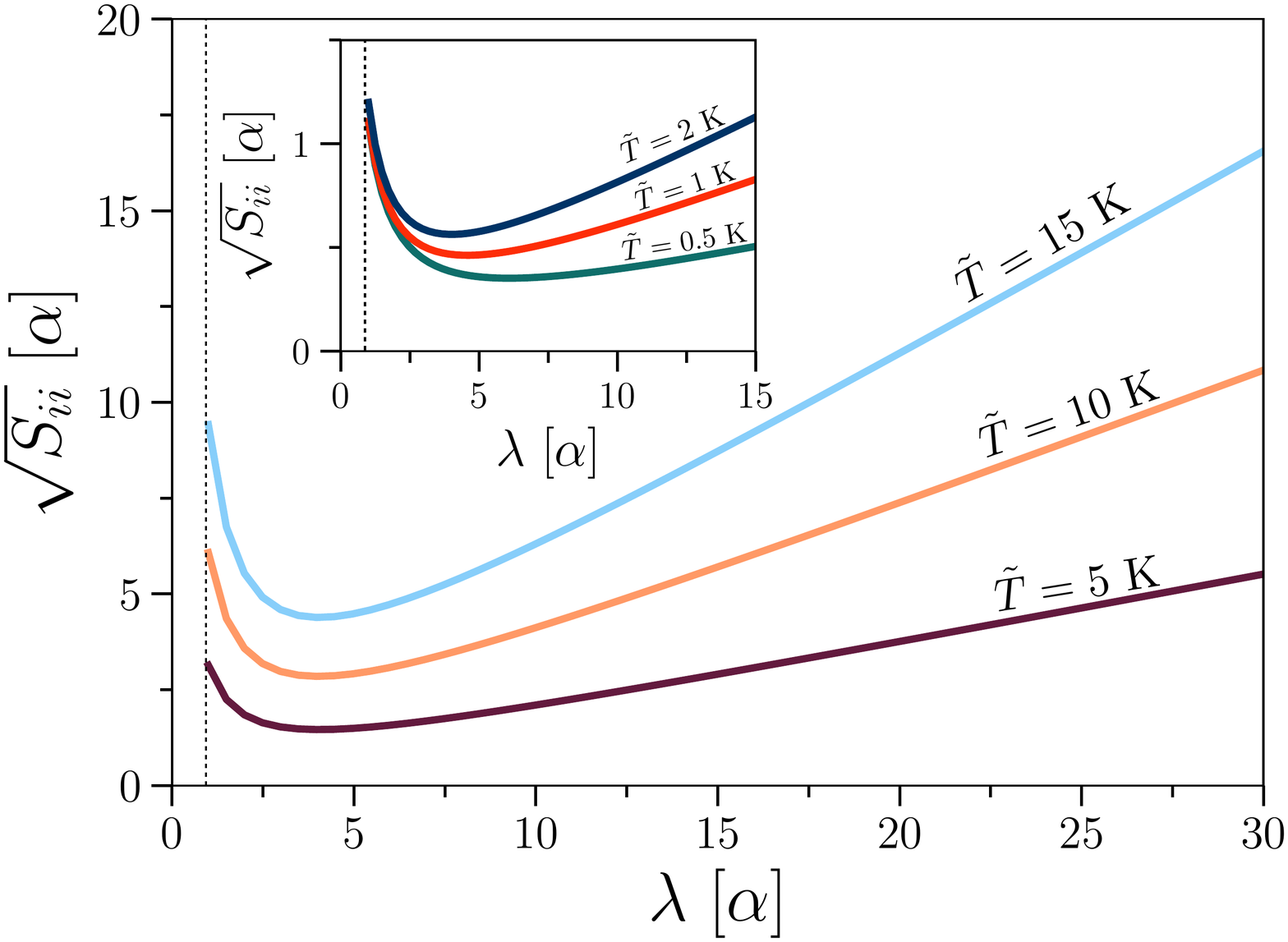}
\caption{The RMSD $\sqrt{S_{ii}}$ given in Eq.~\eqref{MeanSquare} as a function of the skyrmion size $\lambda$ at three different temperatures, $\tilde{T}=5,10,$ and $15$ K, for $J=1$ meV, $t=50$ ps, and $N_AS=1$. The dashed vertical line indicates the value $\lambda =\alpha$. The inset depicts the $\lambda$-dependence of the RMSD at low temperatures below $2$ K. For a given temperature $\tilde{T}$, the RMSD has a local minimum at a critical radius $\lambda_{cr} (\tilde{T})$ which signals a crossover  from   short-time dynamical effects to long-time renormalization: For $\lambda < \lambda_{cr}$, the RMSD  decreases as $1/\lambda$, while for $\lambda > \lambda_{cr}$ it scales linearly with $\lambda$.}
 \label{sTvarious}
\end{figure}
%%%%%%%%%%%%%%%%%%%%%%%%%%%%%%%%%%%%%%%%%%%%%%%%%%%

First we focus on the temperature dependence of the root mean square displacement (RMSD) $\sqrt{S_{ii}(\bar{t})}$, which is summarized in Fig.~\ref{sT}. As a result of the quantum magnetic excitations, the RMSD at  $\tilde{T}=0$, defined as $S_Q = \sqrt{S_{ii}}(\tilde{T}=0)$, remains finite. The dependence of $S_Q$ on the skyrmion size $\lambda$, illustrated in the inset of Fig.~\ref{sT}, implies that quantum fluctuations become important for very small skyrmions of a few lattice sites, while their effect on the RMSD becomes negligible for larger skyrmions. We should emphasize that in this work we consider a classical skyrmion coupled to a bath of quantum magnetic excitations, and disregard quantum effects of the center-of-mass, which could increase the value of $S_Q$ further and make it experimentally more accessible. Such quantum effects are beyond the scope of this paper, and we leave it as a motivation for further studies. 

Another important feature of Fig.~\ref{sT} is the fast linear thermal activation for temperatures $\tilde{T}> 4$ K, i.e., $\sqrt{S_{ii}(\bar{t})} \simeq 0.14 \tilde{T}$ $\alpha$/K. Such a behavior results from the nontrivial temperature dependence of the fluctuation-dissipation theorem Eq.~\eqref{powerspec} and stands in contrast to the $\sqrt{T}$ dependence obtained in a classical description \cite{Miltat18}. For a skyrmion with a radius $10 \alpha$, the RMSD is $(3.3/N_A S)$ percentage of its radius at $\tilde{T}=1.5$ K,  $(10.8/N_A S) $ percentage at $\tilde{T}=5$ K, and $(32.5/N_A S) $ percentage at $\tilde{T}=15$ K. 

Further results are shown in Fig.~\ref{sTvarious}, where we plot the dependence of the RMSD on the skyrmion size $\lambda$. We note that there is a critical radius $\lambda_{cr}(\tilde{T})$ which signals the interplay between long-time renormalization and short-time dynamical effects. For $\lambda<\lambda_{cr}$, the RMSD is inversely proportional to the skyrmion size, as expected for a massive particle with a mass proportional to the area $\lambda^2$. Indeed, the time-dependent damping kernel $\gamma^0_{ij}(t)$ of Eq.~\eqref{KernelDiag0} is renormalized to the effective mass of Eq.~\eqref{Mass0} in the long-time scale approximation. On the contrary, for $\lambda > \lambda_{cr}$, shorter time scale dynamical information becomes dominant and the RMSD scales linearly with $\lambda$. Analogous results are obtained for very low temperatures below $2$ K, illustrated in the inset of Fig.~\ref{sTvarious}.

%%%%%%%%%%%%%%%%%%%%%%%%%%%%%%%%%%%%%%%%%%%%%%%%%%%
\begin{figure}[t]
\centering
\includegraphics[width=1\linewidth]{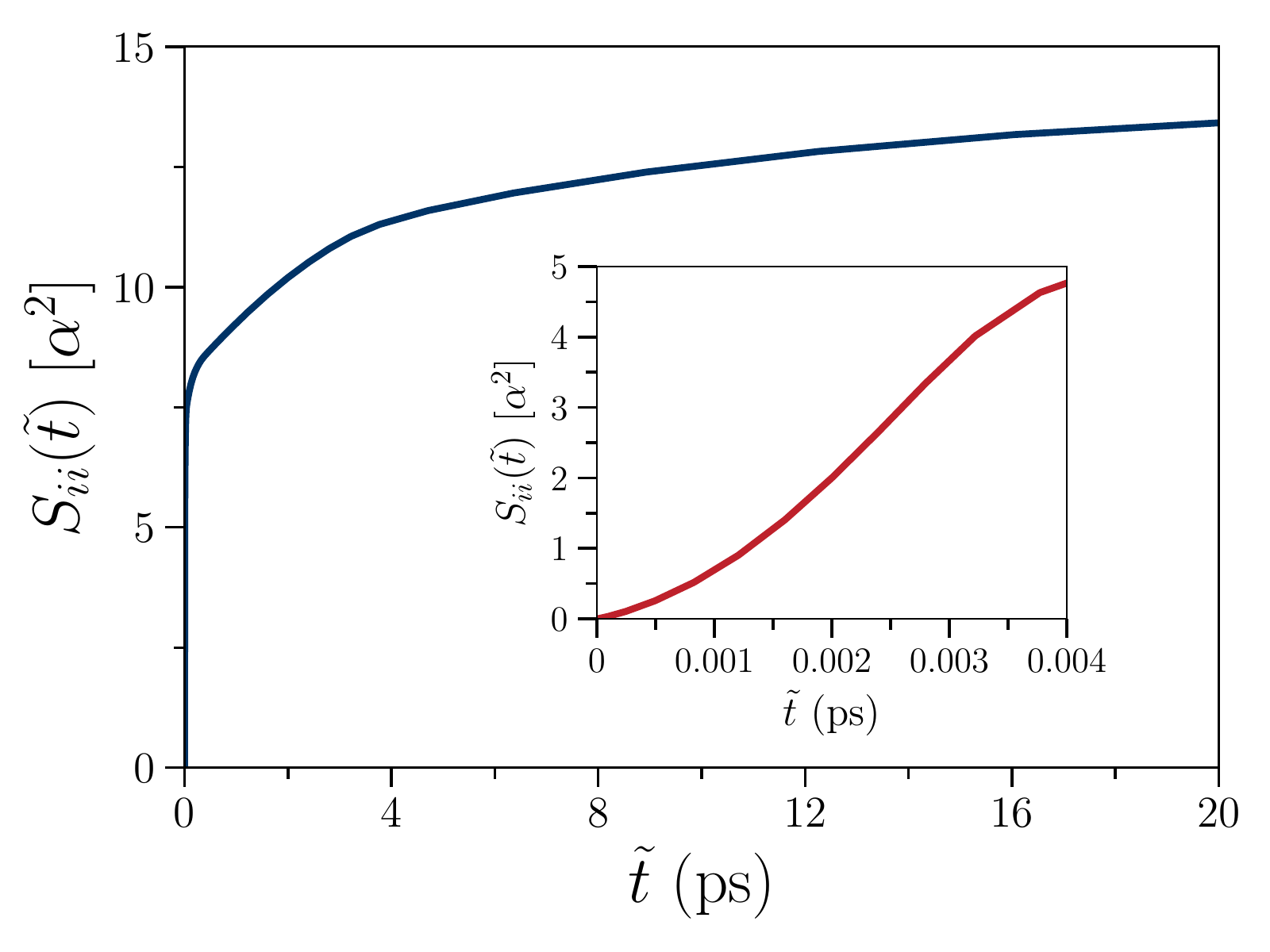}
\caption{Mean squared displacement (MSD) $S_{ii}(\tilde{t})$ given in Eq.~\eqref{MeanSquare} as a function of time at temperature $\tilde{T}=14.5$ K, for a skyrmion of radius $\lambda=3.63 \alpha$. The MSD is plotted for the choice of $J=1$ meV, $N_A S=1$, and $d =4$. We observe a ballistic regime at a very small time-scale, while for larger times the MSD saturates quickly at the value obtained when the memory effects become negligible.}
 \label{Sii}
\end{figure}
%%%%%%%%%%%%%%%%%%%%%%%%%%%%%%%%%%%%%%%%%%%%%%%%%%%

Several conclusions can be drawn also from the time dependence of $S_{ii}(\bar{t})$ as illustrated in Fig.~\ref{Sii}. At short times $\bar{t} \ll 1$, we find a quadratic dependence, $S_{ii}(\bar{t}) \simeq S_0 \bar{t}^2$, which resembles the ballistic regime of the Brownian motion of a particle \cite{Li2013}. The constant $S_0$ is found from Eq.~\ref{MeanSquare} under the replacement $\sin[(\varepsilon_{\nu'}-\varepsilon_{\nu}) \bar{t}/2] \rightarrow  (\varepsilon_{\nu'}-\varepsilon_{\nu}) \bar{t}/2$, while for the specific parameters plotted in Fig.~\ref{Sii} we find $S_0= 4.2 \times 10^5$. Such a ballistic motion is a direct consequence of the memory effects which dominate the dynamics at short time scales. At longer times, the memory effects become negligible and $S_{ii}(\bar{t})$ saturates at a value which can be estimated from replacing $\sin^2[(\varepsilon_{\nu'}-\varepsilon_{\nu}) \bar{t}/2] \rightarrow 1/2$ in Eq.~\eqref{MeanSquare}. 

The ballistic regime of the Brownian motion for a classical particle with a large inertia mass of the order of $10^{-14}$ kg has been experimentally observed for short time scales of the inertia-dominated regime of $\mu$s \cite{Li2010,Huang2011}. Here, the ballistic motion we predict for the quantum dymanics of a magnetic skyrmion, with an inertial mass of $0.2\times 10^{-28}$ kg at $T=580$ mK, is restricted to the immeasurably small femtosecond regime, which, however, is comparable to the duration of ultrafast light-induced heat pulses needed to  write and erase magnetic skyrmions \cite{Berruto2018}. We anticipate that the ballistic motion for a confined skyrmion with an inertial mass of about $10^{-26}$ kg \cite{Psaroudaki17}, could possibly take place within the experimentally accessible nanosecond regime. It suffices to mention that the classical dissipation is dominated by the contribution of some low-lying localized modes with energy $\varepsilon_0$ in the GHz regime \cite{Kravchuk18}. Thus, the quadratic short-time expansion is valid up to times $\varepsilon_0^{-1}$, i.e. the ballistic regime extends in the nanosecond regime. We also note that our predictions significantly deviate from the classical results for the mean squared displacement which, in the latter case, increases linearly with time \cite{Miltat18}, a result that directly follows from the assumption of a phenomenological thermal white noise which scales proportional to the Gilbert damping parameter.

%%%%%%%%%%%%%%%%%%%%%%%%%%%%%%%%%%%%%%%%%%%%%%%%%%%%%%%%%%%%%%%%%%%%%%%%
%%                           Conclusion
%%
%%%%%%%%%%%%%%%%%%%%%%%%%%%%%%%%%%%%%%%%%%%%%%%%%%%%%%%%%%%%%%%%%%%%%%%%
\section{Conclusions}\label{sec:Conclusions} 
In this work, we consider the stochastic dynamics of a magnetic skyrmion in contact with a dissipative bath of magnons in the presence of a time-periodic external field, which directly couples to the magnon bath. We develop a microscopic derivation of the Langevin equation of motion based on a quantum field theory approach which combines the functional Keldysh and the collective coordinate formalism. The non-Markovian damping kernel is explicitly related to the colored autocorrelation function of the stochastic fluctuating fields, through the quantum mechanical version of the fluctuation-dissipation theorem. Emphasis is given to the nontrivial temperature dependence of the dynamical properties of the system, in terms of the fundamental response and correlation functions. Contrary to the prediction of the classical theory, the damping kernel and the mass remain finite at vanishingly small temperatures, due to the quantum nature of the bath considered in this work. This will give rise to a finite mean squared displacement at $T \rightarrow 0$, which increases with temperature as $T^{2}$, a result that deviates from the phenomenological prediction of a linear increase. 

We rigorously treat the effects of an external drive on the bath, and therefore on the skyrmion-bath coupling, and we generalize the theory of quantum dissipative response. The bath is dynamically engineered out-of-equilibrium and through its interaction with the skyrmion gives rise to dissipation and random forces that incorporate the bath’s dynamical activity. The magnitude of these effects is illustrated in the diagonal and off-diagonal response functions, which acquire an additional time-periodicity inherited by the external drive. In addition, a super-Ohmic to Ohmic crossover behavior is signalled by new friction and topological charge renormalization terms, similar to the effects predicted within a microscopic theory of classical dissipation with measurable consequences for the skyrmion path \cite{Psaroudaki18}. We note, however, that, in contrast to Ref.~\onlinecite{Psaroudaki18}, where the external drive couples to a well-pronounced bath mode, here we do not consider resonance effects. 

Within our path integral formulation, we are able to establish a generalization of the fluctuation-dissipation theorem to the nonequilibrium case for weakly driven magnetic excitations. The spectral characteristics of the bath modes of the damping kernel are related to the ones of the stochastic correlation function, irrespectively of the form of the external drive. Noteworthy, our results apply to similar mesoscopic systems embedded in an driven bath. Advances in the theoretical understanding of skyrmion dynamics out of equilibrium is expected to have an impact on similar particle-like objects such as solitonic textures in quantum superfluids and domain walls in ferromagnets. Our  nonequilibrium formalism of skyrmion dynamics can serve as a basis for future experimental investigations as well as theoretical studies that go beyond first order perturbation theory and beyond the slow dynamics of the GHz regime. 

\section{ACKNOWLEDGMENTS} This work was supported by the Swiss National Science Foundation (Switzerland) and the NCCR QSIT.
\appendix
\section{Autocorrelation Function} \label{App:AutoCorr}
Here, we present in more detail the autocorrelation function of the stochastic fields, $C_{ij}(t,t')$, defined in Eq.~\eqref{ActionSt} of Sec.~\ref{sec:Model}. To evaluate the trace we use the functions $\Psi_{\nu}(\mb{r},t)$, eigenfunctions of the magnon Hamiltonian $\mc{H}$, which are presented in detail in Appendix~\ref{App:Magnons}. After some algebra, $C_{ij}(t,t')$ is expressed as,
\begin{align}
C_{ij}(t,t')= C_{ij}^{K,K}(t,t') + C_{ij}^{R,A}(t,t') + C_{ij}^{A,R}(t,t') \,,
\end{align}
where 
%the resulting expressions for functions $C_{ij}^{a,b}(t,t')$ are
\begin{align}
C_{ij}^{a,b}(t,t')= \frac{i S^2}{4} \sum_{\nu}\!^{}{'}\,\, \int_{\bar{t}} \int_{\bar{\mb{r}},\mb{r},\mb{r}'} \partial_{t}\partial_{t'} \left[\Psi_{\nu}^{\dagger}(\bar{\mb{r}},\bar{t}) G^{a}(\bar{\mb{r}},\mb{r}',\bar{t},t')  \right. \nonumber \\ 
\left. \times \Gamma_i(\mb{r}') \sigma_z  G^{b}(\mb{r}',\mb{r},t',t) \sigma_z \Gamma_j(\mb{r}) \Psi_{\nu}(\mb{r},t)  \right]\,,
\label{CorrelatorsBC}
\end{align}
with $a,b = K,R,A$. 	A more transparent form is obtained for a bath of magnetic excitations at equilibrium, i.e. $\mb{b}(t)=0$, 
\begin{align}
C^0_{ij}(t-t') &= \frac{i}{4} \pt_{t}\pt_{t'}\sum_{\nu,\nu'}\!^{}{'}\,\, B_{ij}^{\nu\nu'}e^{-i (\varepsilon_{\nu'} -\varepsilon_{\nu})(t-t')} \nonumber \\ 
&\times [\Theta(t-t')+\Theta(t'-t)-F(\varepsilon_{\nu'})F(\varepsilon_{\nu})],
\label{C0t}
\end{align} 
which is further simplified in Fourier space,
\begin{equation}
C^0_{ij}(\omega) = \frac{i \pi\omega^2 }{2}\coth(\frac{\beta \omega}{2}) \sum_{\nu,\nu'}\!^{}{'}\,\, B_{ij}^{\nu\nu'}\bar{F}_{\nu'\nu}\delta(\omega - \varepsilon_{\nu'} +\varepsilon_{\nu}),
\end{equation} 
where, again, $\bar{F}_{\nu\nu'} = F(\varepsilon_{\nu})-F(\varepsilon_{\nu'})$ and $F(\varepsilon_{\nu})=\coth(\beta \varepsilon_{\nu}/2)$. 

\section{Equilibrium Damping Kernel}\label{App:EqDiss}
Our current task is to analyze the damping kernel of Eq.~\eqref{FrictionKernel} by considering first the special case $\mb{b}(t)=0$. By a simple inspection of Eq.~\eqref{GreenFun} we notice that correlations in equilibrium are time translation invariant and the Green functions depend on time differences, $G^{R,A}(t,t') = G^{R,A}(t-t')$ and as a result the the diagonal part of the damping kernel is found equal to 
\begin{equation}
\gamma^{0}_{ii}(t) =\Theta(t) \partial_t \sum_{\nu\nu'}\!^{}{'}\,\, \Re(\mc{B}_{ii}^{\nu\nu'}) \bar{F}_{\nu\nu'}\sin [(\varepsilon_{\nu'} -\varepsilon_{\nu})t] \,,
\label{KernelDiagTime}
\end{equation}
while the off-diagonal part can be cast into the form 
\begin{equation}
\gamma^{0}_{yx}(t) =\Theta(t) \partial_t \sum_{\nu,\nu'}\!^{}{'}\,\, \Im(\mc{B}_{yx}^{\nu\nu'}) \bar{F}_{\nu\nu'}\cos [(\varepsilon_{\nu'} -\varepsilon_{\nu})t] \,.
\label{KerneloffDiagTime}
\end{equation}
Here we sum over the quantum number $\nu = \{q=\pm 1,n\}$, where the index $q$ distinguishes between particle states ($q=1$), solutions of the eigenvalue problem $\mc{H} \Psi_n = \varepsilon^{q}_n \sigma_z \Psi_n$, with positive eigenfrequency $\varepsilon_n^{1} =+ \varepsilon_n$, and antiparticle states ($q=-1$) with negative eigenfrequency $\varepsilon_n^{-1} =- \varepsilon_n$ \cite{Psaroudaki17}. The matrix elements are given by $\mc{B}_{ij}^{\nu\nu'} = \mc{B}_{ij}^{n,q;n',q'} =  (q q'/2)  \int_{\mb{r}} \Psi_{\nu}^{\dagger} \Gamma_i \sigma_z\Psi_{\nu'} \int_{\mb{r'}} \Psi_{\nu'}^{\dagger} \Gamma_j \sigma_z \Psi_{\nu}$.
%$\bar{F}_{\nu\nu'} = [F(\varepsilon_{\nu})-F(\varepsilon_{\nu'})]$, and $F(\varepsilon_{\nu})=\coth(\beta \varepsilon_{\nu}/2)$. 
From the structure of the matrix elements we conclude that $\mc{B}_{ii}^{\nu\nu'} = \Re(\mc{B}_{ii}^{\nu\nu'})$ and $\mc{B}_{xy}^{\nu\nu'} = i \Im(\mc{B}_{ij}^{\nu\nu'})$. We also note that $\gamma^0_{xy}(t)= -\gamma^0_{yx}(t)$ and that $\Re(\mc{B}_{ii}^{\nu'\nu})=\Re(\mc{B}_{ii}^{\nu\nu'})$, while $\Im(\mc{B}_{yx}^{\nu'\nu})= -\Im(\mc{B}_{yx}^{\nu\nu'})$. Thus, both Eqs.~\eqref{KernelDiagTime} and \eqref{KerneloffDiagTime} are symmetric under the interchange of  $\nu$ and $\nu'$. 

It appears convenient to derive the Langevin equation of Eq.~\eqref{EquationTime} in the Laplace-frequency $z$ space,
\begin{align}
\tilde{Q}_0 \epsilon_{ij} z  R_c^{j}(z)  +z R_c^{j}(z) \gamma_{ji}(z) =\xi_i(z) \,,
\label{EqFreq}
\end{align}
where the frequency dependent kernel of Eq.~\eqref{KernelDiagTime} equals
\begin{align}
\gamma^{0}_{ii}(z) = z \sum_{\nu,\nu'}\!^{}{'}\,\, \frac{\Re(\mc{B}_{ii}^{\nu;\nu'}) (\varepsilon_{\nu'} -\varepsilon_{\nu}) \bar{F}_{\nu\nu'}}{(\varepsilon_{\nu'}-\varepsilon_{\nu})^2+z^2} \,,
\label{KernelDiag0}
\end{align}
and the off-diagonal kernel of Eq.~\eqref{KerneloffDiagTime} is found to be 
\begin{align}
\gamma^{0}_{xy}(z) =z^2 \sum_{\nu,\nu'}\!^{}{'}\,\, \frac{\Im[\mc{B}_{xy}^{\nu;\nu'}]\bar{F}_{\nu\nu'}}{(\varepsilon_{\nu'}-\varepsilon_{\nu})^2+z^2} \,,
\label{KernelOff0}
\end{align}
and it also holds that $\gamma^{0}_{yx}(z)= -\gamma^{0}_{xy}(z)$. Note that agreement of the damping kernel $\gamma_{ij}^{0}(z)$ with earlier results derived in Matsubara space using the imaginary time path integral approach \cite{Psaroudaki17} can be established by simple analytic continuation. 

\section{Nonequilibrium Damping Kernel}\label{App:NonEqDiss}
In this section we provide explicit formulas for the reduced expressions of the nonequilibrium damping kernels appearing in Eq.~\eqref{Diagonal}. First we note that the Laplace transform $W_{ij}(z)$ of the function $W_{ij}(t)$ given in Eq.~\eqref{DampingTimeCor} is expressed as
\begin{equation}
W_{ji}(z) =\sum_{\nu_1, \nu_2,\nu_2}\!\!\!\!^{}{'}\,\,\frac{\mc{C}_{ji}^{\nu_1\nu_2\nu_3}[ w_{\nu_3\nu_2}(z)-w_{\nu_3\nu_1}(z)]}{(\varepsilon_{\nu_2}- \varepsilon_{\nu_1})^2 -\wext^2}  \,,
\label{DampingTimeCorZ}
\end{equation}
where $w_{\nu_1\nu_2}(z) = \bar{F}_{\nu_1\nu_2} (\varepsilon_{\nu_1}-\varepsilon_{\nu_2})/[(\varepsilon_{\nu_1}-\varepsilon_{\nu_2})^2+z^2]$ and where the matrix elements $\mc{C}_{ji}^{\nu_1\nu_2\nu_3}$ are given in Eq.~\eqref{MatrixEl}. Starting from Eq.~\eqref{NonEqDissi} we define the temperature dependent dissipation constants through a Taylor expansion of the kernel $\Delta \gamma_{ji}(t,z)$ around  $z=0$ as
\begin{equation}
 \Delta \gamma_{xx}(t,0) = -\wext [W_{xx}(0)+W_{xx}(i \wext) ] \sin(\wext t)
 \end{equation}
for the linear friction-like terms and the next order is given through the relation 
\begin{align}
\pt_{z} \Delta \gamma_{xx}(t,z)\vert_{z=0}&=\cos(\wext t)[W_{xx}(0)+W_{xx}(i \wext)\nonumber \\&+ i \wext W_{xx}'(i \wext)] \,.  
\end{align}
Analogously, we find 
\begin{align}
\Delta \gamma_{yx}(t,0)&= \wext [W_{yx}(0)+W_{yx}(i \wext) ] \cos(\wext t) 
\end{align}
 for the linear friction-like terms and 
 \begin{align}
 \pt_{z} \Delta \gamma_{yx}(t,z)\vert_{z=0}&= \sin(\wext t)[ W_{yx}(0)+W_{yx}(i \wext)\nonumber \\&+ i \wext W_{yx}'(i \wext)] \,.  
 \end{align}
for the next order term. First we note that in the static limit $\wext \rightarrow 0$, all the terms vanish besides a mass renormalization term  $W_{xx}(0)$. At this point we emphasize that our approach is valid only for slow dynamics and consequently the frequency of the external drive should be restricted to the GHz range, i.e. $\wext \ll \egap$. At the same time we recall that the external potential $V$ induces a finite but small overlap $0< \vert V_{\nu_1\nu_2} \vert \ll 1$ between magnon modes carrying approximately the same energy. Thus, under the assumptions  $\varepsilon_d \ll \wext \ll \egap$, with $\varepsilon_d=\vert \varepsilon_{\nu_1}- \varepsilon_{\nu_2} \vert$, the resulting expressions are summarized in Eqs.~\eqref{Diagonal}--\eqref{offDiagonal}, with $D(T) = -\wext \bar{W}_{ii}$, $\delta M(T) = \bar{W}_{ii}$, $\delta Q(T) = \wext \bar{W}_{yx}$ and $G(T) = \bar{W}_{yx}$. The $\bar{W}_{ji}$ coefficient can be found in Eq.~\eqref{W0}.

\section{Magnon Spectrum}\label{App:Magnons}
Here we briefly discuss the structure of the magnon excitations, while a more detailed discussion can be found in Refs. \onlinecite{Psaroudaki17, Psaroudaki18}. The magnon Hamiltonian $\mc{H}$ for the model of Eq.~\eqref{FreeEnergy} in dimensionless units is given by
\begin{equation}
\mc{H}=2[- \nabla^2 + U_0(\rho)] \mathds{1} + 2 U_1(\rho) \sigma_x - 2 i U_2(\rho) \frac{\pt}{\pt \phi}\sigma_z \,,
\label{EVPTransf}
\end{equation}
where $U_2(\rho)=\frac{2\cos \Theta_0}{\rho^2} -\frac{ \sin \Theta_0}{\rho}$, and 
\begin{align}
U_1(\rho)&=\frac{\sin 2\Theta_0 }{4\rho}- \frac{(\Theta_0')^2}{2}+ \frac{1}{2}(\kappa+\frac{1}{\rho^2})\sin^2 \Theta_0 - \frac{ \Theta_0'}{2} \,,
\end{align}
and 
\begin{align}
U_0(\rho)&= \frac{h \cos \Theta_0}{2}  -\frac{3\sin 2 \Theta_0 }{4 \rho}  \nonumber \\
& -\frac{(\Theta_0')^2}{2}+(\frac{\kappa}{4}+\frac{1}{4 \rho^2} )(1+ 3 \cos 2\Theta_0)-\frac{\Theta_0'}{2} \,.  
\end{align}

The goal is to solve the eigenvalue problem of the form $\mc{H} \Psi_{n} =  \varepsilon_{n} \sigma_z \Psi_{n}$. Using the wave expansions $\Psi_{n}=e^{i  m \phi}\psi_{n,m}(\rho)/\sqrt{2 \pi}$, the eigenvalue problem takes the form $\mc{H}_m  \psi_{n,m}(\rho) = \varepsilon_{n,m} \sigma_z  \psi_{n,m}(\rho)$, with 
\begin{equation}
\mc{H}_m =  2(- \nabla^2_{\rho}+ U_0(\rho) +\frac{ m^2}{\rho^2} ) \mathds{1}  
+ 2U_1(\rho) \sigma_x+ 2U_2(\rho) m \sigma_z\,,
\label{EVPTransfm}
\end{equation}
and $\nabla^2_{\rho}= \frac{\pt^2}{\pt_{\rho}}+\frac{1}{\rho} \frac{\pt}{\pt_{\rho}}$. Scattering states $\Psi_{m,k}(\mb{r})$, classified by $m$ as well as the radial momentum $k \geqslant 0$, carry energy $\varepsilon(k) = \egap + k^2$, with $\egap = 2\kappa + h$, and are of the form
\begin{equation}
\psi_{m,k}(\rho)= d_m \left[\cos(\delta_m) J_{m+1}(k \rho)- \sin(\delta_m) Y_{m+1}(k \rho) \right] \binom{1}{0}\,,
\label{SC}
\end{equation}
where $J_m~(Y_m)$ are the Bessel functions of the first (second) kind, $d_m(k)$ is a normalization constant and $\delta_m(k)$ is a scattering phase shift that determines the intensity of magnon scattering due to the presence of the skyrmion. The phase shifts are  calculated within the WKB approximation discussed in detail in Refs.~\onlinecite{Psaroudaki17, Berry72}. In the presence of an oscillating field $\mb{b}(t)= b_0 \Theta(t-t_0) \cos(\wext t ) ( \sin \phiext,0, \cos \phiext)$,  the magnons experience a potential $V(\mb{r},t) = \mb{b}(t) \cdot \mb{D} = b_0 \Theta(t-t_0) \cos(\wext t) V(\mb{r})$, with $ \mb{D} =  \delta_{\chi^\dagger}\delta_\chi \mb{m} |_{\chi=\chi^\dagger=0}$ and 
\begin{equation}
V(\mb{r}) =V_1(\mb{r}) \mathds{1} + V_2(\mb{r}) \sigma_x + V_3(\mb{r}) \sigma_y \,.
\label{PotentialDrive}
\end{equation}
The potentials are $V_{1,2}(\mb{r})= \csc [\Theta_0(\rho)]^2 B_1(\mb{r})\pm B_2 (\mb{r})$, $V_3(\mb{r}) = (1/2) \csc [\Theta_0(\rho)]  B_3(\mb{r})+ B_2 (\mb{r})$, and $B_1(\mb{r})=(1/2) \sin(\phiext) \cos[\Phi_0(\phi)] \sin [\Theta_0(\rho)]$, $B_2(\mb{r})= B_1(\mb{r}) + (1/2) \cos(\phiext) \cos[\Phi_0(\phi)] \sin[\Theta(\rho)]$ and $B_1(\mb{r})= -(1/2) \sin(\phiext) \sin[\Phi_0(\phi) ]\cos [\Theta_0(\rho)]$.

We note that since the Hamiltonian $\mc{H}$ is invariant under the  conjugation transformation $\mc{C}$, where $\mc{C}=K \sigma_x$ with $K$ the complex conjugation operator, there exists an additional class of solutions $\Psi_n^{-1}= \mc{C}\sigma_x \Psi^1_{n} $ with negative eigenfrequency. To distinguish these two classes of solutions we use an additional index $\Psi^{q=\pm1}_{n}$, where the states $\Psi^{1}_{n}$ have positive eigenfrequencies $\varepsilon^{1}_n \geq 0$, while  $\Psi_n^{-1}$ have negative eigenfrequencies $\varepsilon^{-1}_n \leq 0$. The bi-orthogonality conditions for the solutions are of the form $\langle \Psi^{q}_{n} \vert\sigma_z \vert \Psi^{q'}_{m} \rangle = q \delta_{q,q'} \delta_{n,m}$. Similarly, the resolution of the unity operator is given by $\mathds{1} = \sum_{q=\pm 1} \sum_{n} q \vert \Psi^{q}_{n} \rangle \langle \Psi^{q}_{n} \vert \sigma_z$ and the trace of an operator is $\mbox{Tr}(A) = \sum_{q=\pm 1} \sum_{n} q  \langle \Psi^{q}_{n} \vert \sigma_z  A \vert \Psi^{q}_{n} \rangle$. To calculate the mass $\mc{M}(T)$ of Eq.~\eqref{MassT}, as well as the drive-induced dissipation of Eq.~\eqref{W0}, the sum over the quantum number $\nu$ is replaced in the following way: 
\begin{align}
\sum_{\nu} \Psi_\nu = \sum_{q=\pm 1, n} \Psi^{q}_{n}  \rightarrow \sum_{q=\pm 1} \sum_{m} \sum_{k} \Psi^{q}_{m,k} \,.
\end{align}
To render our results finite in the thermodynamic limit, we subtract the background fluctuations \cite{Braun96} as $\sum_{k} \Psi_{m,k} \rightarrow \sum_k \left(\Pfr_{m,k}  -\Psi_{m,k}\right)$, where $\Pfr_{m,k}$ are given by Eq.~\eqref{SC} for $\delta_m(k) =0$. 
We also note that in addition to scattering states, a few localized modes which correspond to deformations of the skyrmion into polygons exist in the range $0 < \varepsilon_n < \egap$, but do not contribute significantly compared to the continuum of modes $\Psi_{m,k}$. For detailed formulas of the explicit calculation of the mass $\mc{M}(T)$ we refer the reader to Appendix C  of Ref.~\onlinecite{Psaroudaki17} and in particular to Eq.~(C8).

\end{document}